\documentclass{article}
\usepackage{arxiv}

\usepackage[utf8]{inputenc} 
\usepackage[T1]{fontenc}    
\usepackage{hyperref}       
\usepackage{url}            
\usepackage{booktabs}       
\usepackage{amsfonts}       
\usepackage{nicefrac}       
\usepackage{microtype}      
\usepackage{lipsum}
\usepackage{float}
\usepackage{graphicx}
\usepackage{multirow}
\usepackage{array}
\usepackage{comment}
\usepackage{longtable}
\usepackage{ragged2e}
\usepackage{subfig}
\usepackage[table]{xcolor}
\usepackage{pgfplots}
\usepackage{geometry}
\usepackage{multicol}
\usepackage{amsmath}
\usepackage{pifont}
\pgfplotsset{compat=1.18}
\title{Artificial Intelligence and Journalism: A Systematic Bibliometric and Thematic Analysis of Global Research}

 \author{
    Mohammad Al Masum Molla \\
    Gaylord College of Journalism and Mass Communication\\
    University of Oklahoma\\
    Norman, Oklahoma-73019\\
    \texttt{mohammadmasum@ou.edu} \\
    \And
    Md Manjurul Ahsan \\
    Department of Industrial and Systems Engineering \\
    University of Oklahoma \\
    Norman, Oklahoma-73071 \\
    \texttt{ahsan@ou.edu} \\
}

\begin{document}
\maketitle

\begin{abstract}
Artificial Intelligence (AI) is reshaping journalistic practices across the globe, offering new opportunities while raising ethical, professional, and societal concerns. This study presents a comprehensive systematic review of published articles on AI in journalism from 2010 to 2025. Following the Preferred Reporting Items for Systematic Reviews and Meta-Analyses (PRISMA) 2020 guidelines, a total of 72 peer-reviewed articles were selected from Scopus and Web of Science databases. The analysis combines bibliometric mapping and qualitative thematic synthesis to identify dominant trends, technologies, geographical distributions, and ethical debates. Additionally, sentiment analysis was performed on article abstracts using the Valence Aware Dictionary and sEntiment Reasoner (VADER) algorithm to capture evaluative tones across the literature. The findings show a sharp increase in research activity after 2020, with prominent focus areas including automation, misinformation, and ethical governance. While most studies reflect cautious optimism, concerns over bias, transparency, and accountability remain persistent. The review also highlights regional disparities in scholarly contributions, with limited representation from the Global South. By integrating quantitative and qualitative insights, this study offers a multi-dimensional understanding of how AI is transforming journalism and proposes future research directions for inclusive and responsible innovation.

\end{abstract}



\section{Introduction}\label{sec:intro}

Artificial intelligence (AI) is increasingly integrated into journalistic practice, spanning automation, natural-language processing (NLP), computer vision, and, most recently, generative models such as ChatGPT, Claude, and Gemini\,\cite{brennen2020,graefe2016,torres2023,adjin2024role,aleessawi2024implications,Almalki2022Incorporating,Almania2024Mechanisms}.  
These technologies now support every stage of news production—from content creation and verification to distribution and audience engagement\,\cite{graefe2016,FieirasCeide2024AI,garcia2020robots,Gherhes2024Are}.  
While early automation handled routine, data-driven beats (e.g.\ sports or finance), recent generative breakthroughs enable entire stories, summaries, headlines, and visuals with minimal human input\,\cite{brennen2020}.

The late-2022 release of ChatGPT proved a watershed: within months it surpassed 100 million users, amplifying both experimentation and anxiety in newsrooms\,\cite{doembana2025,buchholz2023,fomenko2024,Gherhes2024Are,Huh2025Can,LermannHenestrosa2023Automated,Schaetz2025AI}.  
Large language models (LLMs) now deliver context-aware text that mimics human prose. Major organisations—including the Associated Press, \emph{Washington Post}, and BBC—use AI for data analysis, productivity, and audience engagement\,\cite{thurman2019,fieiras2024,FieirasCeide2024AI}.  
Yet promises of efficiency collide with ethical worries over misinformation, bias, opacity, and diminished editorial control\,\cite{mahony2024,molitorisz2024,Chen2023Gendered,Gonzalez-Arias2024Rethinking,Johanssen2021Artificial,Kuai2022AI,Kuai2025Navigating,Matich2025Old}.

AI is therefore not merely a tool but an emerging infrastructure that reshapes journalistic identity and epistemology~\cite{brennen2020,Hermida2024Automata,Matich2025Old,Tejedor2021Exo,Thomas2023What}.  
As automated decision making obscures editorial processes, journalistic authority is renegotiated\,\cite{lewis2025}.  
Four overlapping ``waves’’—recommendation engines, automated text, audience analytics, and generative models—underline this growing complexity\,\cite{torres2023,Thurman2019Algorithms}.  
Public optimism co-exists with professional scepticism, reflecting tensions between innovation and accountability~\cite{gutierrez2023,ErcegMatijasevic2024Assessing,Moran2022Robots,van2024revisiting}.

Despite rising interest since 2016, scholarship remains fragmented conceptually, empirically, and regionally.  
Foundational reviews pre-date the generative-AI wave\,\cite{Apablaza-Campos2024Generative,Goncalves2022Inteligencia,Hassouni2025AI,Ji2024Scrutinizing,Mahony2024Concerns,Mohammed2024Friends,Ncube2025Mind}.  
Most studies emphasise automation or personalisation, with limited focus on how generative AI affects editorial values and public trust\,\cite{doembana2025,pena2023,Gherhes2024Are,Huh2025Can,LermannHenestrosa2023Automated,Schaetz2025AI,Apablaza-Campos2024Generative,sharadga2022journalists}.  
Case studies still skew toward the United States, United Kingdom, and Western Europe, leaving Latin America, Africa, and much of Asia under-represented\,\cite{torres2023,adjin2024role}.  
Interdisciplinary work bridging journalism and computer science is also scarce~\cite{GutierrezLopez2022Question,JannieMoller2024Algorithmic,Johanssen2021Artificial,GutierrezLopez2022Question,JannieMoller2024Algorithmic,Veerbeek2025Fighting,Gutierrez-Caneda2024Ethics}.

Under-explored dimensions include AI’s impact on journalism education, freelancers, small newsrooms, transparency in algorithmic news decisions, and audience perceptions of AI-authored content\,\cite{gondwe2024,gonzalez2024,Ncube2025Mind,Eder2025Falling,Fridman2023How}.  

We conduct a systematic, PRISMA-guided bibliometric review of AI–journalism research published between 2010 and June 2025.  
Using Scopus and Web of Science, we map publication trends, keyword evolution, and collaboration networks.  
Our review asks:
\begin{itemize}
    \item \textbf{RQ1:} What thematic trends define AI-and-journalism scholarship (2010–2025)?
    \item \textbf{RQ2:} How have publication patterns, geographies, and collaboration networks evolved?
    \item \textbf{RQ3:} Which AI technologies and newsroom applications dominate the literature?
    \item \textbf{RQ4:} What ethical, professional, and social concerns recur?
\end{itemize}
The review consolidates fragmented insights into a coherent overview of a critical 15-year window, flagging regional gaps, under-studied technologies (e.g.\ generative visuals), and neglected topics such as long-term newsroom change and journalism education.  
Findings guide scholars, practitioners, and policymakers toward more inclusive, accountable, and transparent AI adoption in journalism.

The rest of the paper is organized as follows: Section~\ref{met} describes the methodology used during this study; Section~\ref{bib} includes the bibliometric analysis of the 274 articles; Section~\ref{sen} presents the sentiment analysis of the referenced literature using Natural Language Processing; Section~\ref{ai} provides a literature review and general insights into AI in journalism; Section~\ref{dis} discusses the overall findings; and finally, an overall conclusion is drawn in Section~\ref{con}.
\section{Methodology}\label{met}

\begin{figure}[h!]
\centering
\includegraphics[width=\textwidth]{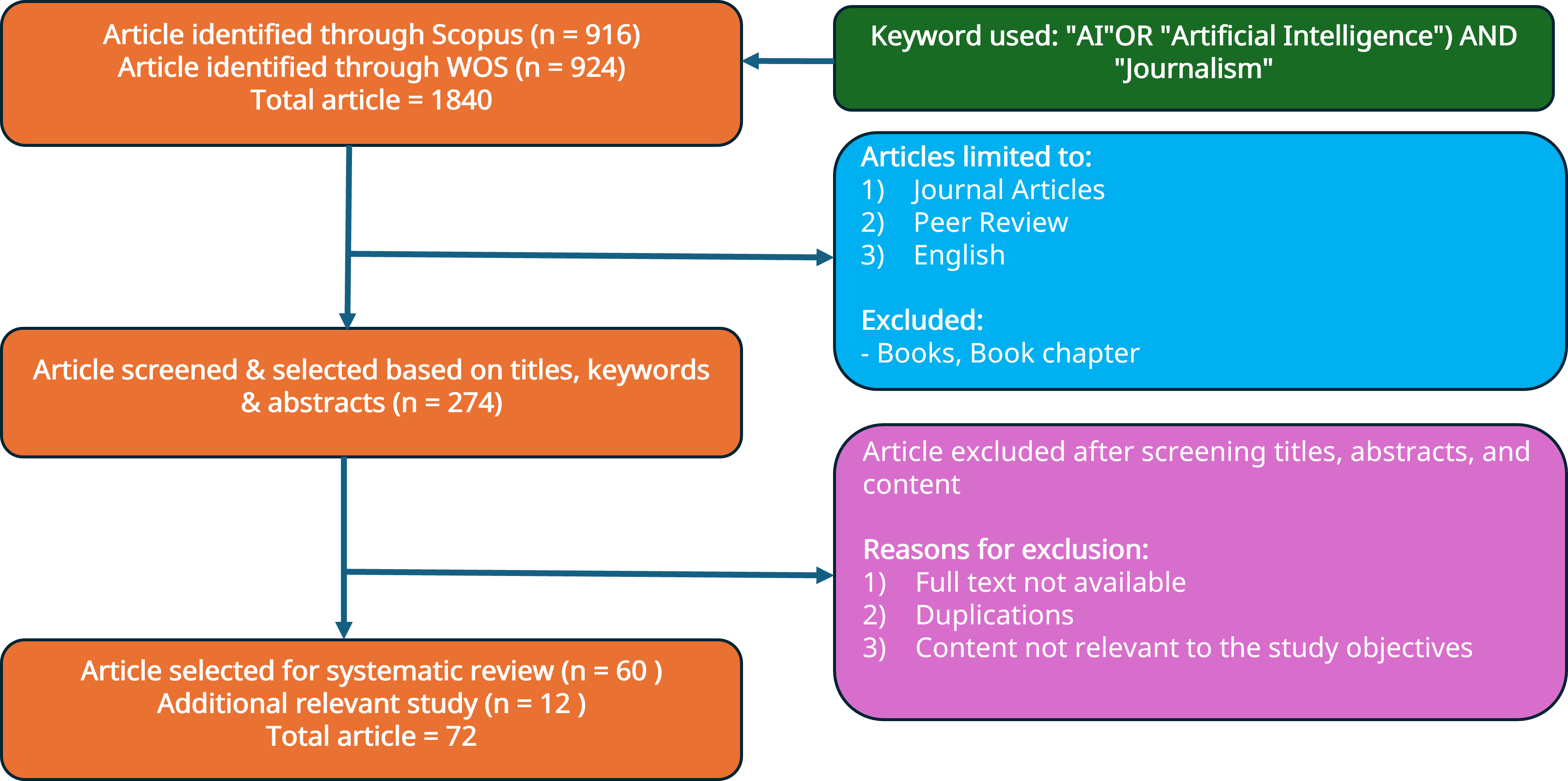}
\caption{Flow diagram of article selection procedure based on PRISMA 2020 guidelines \cite{page2021prisma}.}
\label{fig:prisma}
\end{figure}

\textbf{Systematic Review Protocol.} This study follows the Preferred Reporting Items for Systematic Reviews and Meta-Analyses (PRISMA) 2020 guidelines \cite{page2021prisma} to ensure methodological transparency, reproducibility, and consistency. PRISMA is a widely recognized framework used for conducting systematic literature reviews (SLRs) across disciplines, including media and technology research. The article selection procedure is illustrated in Figure~\ref{fig:prisma}, and detailed inclusion/exclusion criteria are presented in Table~\ref{tab:incl_excl_criteria}.

\textbf{Search Strategy.} A comprehensive database search was conducted on 2 July 2025 using two multidisciplinary academic sources: Elsevier Scopus and Clarivate Web of Science (WoS). The Boolean search string used was \texttt{("AI" OR "Artificial Intelligence") AND "Journalism"}, developed by one investigator and refined in consultation with a second reviewer to ensure both sensitivity and specificity. The search covered the period from 1 January 2010 to 30 June 2025 to capture the contemporary rise and evolution of AI technologies in journalistic practice.

\textbf{Identification and Pre-screening.} The initial search retrieved 916 articles from Scopus and 924 from WoS, totaling 1,840 records. After applying filters for peer-reviewed journal articles, English language, and open-access availability, 352 records remained. These were exported into EndNote, and duplicates were removed using Excel’s duplication tools \cite{ahsan2022machine,mustapha2021impact,ahsan2022machinea}.

\textbf{Screening and Eligibility.} The remaining 274 unique articles were independently screened at the title and abstract level by two reviewers (M.A. and M.M.) based on alignment with the research questions. Discrepancies were resolved through discussion. Articles were included if they met at least one of the following criteria: (i) they examined empirical or conceptual applications of AI in journalism; (ii) they addressed themes such as news production, misinformation, ethics, professional roles, newsroom innovation, or audience engagement; and (iii) they were peer-reviewed and published within the designated timeframe.

During full-text screening, articles were excluded due to: lack of full-text availability, duplication, or irrelevance to the study objectives. An additional 12 relevant studies were identified through backward citation searching. Ultimately, 72 articles were selected for final thematic analysis.

\begin{table}[ht]
\centering
\caption{Inclusion and exclusion procedure of referenced literature in AI and Journalism}
\label{tab:incl_excl_criteria}
\begin{tabular}{@{} l p{8cm} c c @{}}
\toprule
\textbf{Screening Type} & \textbf{Criteria} & \textbf{Included} & \textbf{Excluded} \\
\midrule
Title Screening & Does the title address AI, artificial intelligence, or related terms in journalistic contexts? & 274 & 352 \\
Abstract Screening & Is the abstract focused on AI use, challenges, or impacts in journalism or news media? & 134 & 140 \\
Full-text Screening & Is the full text accessible and does it contain empirical or conceptual discussion on AI in journalism? & 60 & 74 \\
Additional Screening & Was the article located through citation tracking and manually verified for relevance? & 12 & – \\
Total Articles for Final Review & Selected for thematic analysis and qualitative synthesis & 72 & – \\
\bottomrule
\end{tabular}
\end{table}

\subsection{Natural Language Processing and Sentiment Analysis}

To augment our qualitative thematic synthesis, we performed sentiment analysis using Natural Language Processing (NLP) techniques on the abstracts of the selected articles. The aim was to quantify evaluative attitudes toward AI in journalism across time.

Let $\mathcal{D} = \{d_1, d_2, \dots, d_N\}$ denote the collection of $N = 72$ abstracts included in the final review. Each abstract $d_i$ was preprocessed through the following pipeline~\cite{manning1999foundations}:
\begin{itemize}
    \item Lowercasing: $d_i \rightarrow \texttt{lower}(d_i)$
    \item Tokenization: $d_i \rightarrow \{w_1, w_2, \dots, w_k\}$
    \item Stopword removal and lemmatization: $\{w_j\} \rightarrow \{w_j^*\}_{j=1}^k$
\end{itemize}

We applied the VADER sentiment scoring algorithm~\cite{hutto2014vader} to compute the polarity score for each document $d_i$. The compound sentiment score $s_i \in [-1, 1]$ is defined as:
\[
s_i = \text{compound}(d_i) = \frac{\sum_{j=1}^{k} \text{valence}(w_j^*)}{\sqrt{\sum_{j=1}^{k} \text{valence}(w_j^*)^2} + \alpha}
\]
where $\text{valence}(w_j^*)$ is the lexicon-based polarity of word $w_j^*$ and $\alpha$ is a normalization constant empirically set by VADER (default $\alpha = 15$). The score $s_i$ is mapped to three sentiment classes:
\[
\text{sentiment}_i = 
\begin{cases}
\text{Positive} & \text{if } s_i > 0.05 \\
\text{Negative} & \text{if } s_i < -0.05 \\
\text{Neutral} & \text{otherwise}
\end{cases}
\]

Let $P$, $N$, and $Z$ denote the number of documents classified as positive, negative, and neutral, respectively:
\[
P = \sum_{i=1}^{N} \mathbb{1}(s_i > 0.05), \quad
N = \sum_{i=1}^{N} \mathbb{1}(s_i < -0.05), \quad
Z = N - P - N
\]

To explore lexical trends, we constructed polarity-specific term frequency vectors~\cite{liu2012sentiment}:
\[
\textbf{TF}^{(+)} = [f_1^{(+)}, f_2^{(+)}, \dots], \quad
\textbf{TF}^{(-)} = [f_1^{(-)}, f_2^{(-)}, \dots]
\]
where $f_j^{(+)}$ and $f_j^{(-)}$ denote the frequency of the $j$-th most common positive and negative word, respectively, after excluding overlapping entries.

Finally, to assess sentiment evolution over time, we defined the mean annual sentiment $\bar{s}_y$ for year $y$:
\[
\bar{s}_y = \frac{1}{|\mathcal{D}_y|} \sum_{d_i \in \mathcal{D}_y} s_i
\]
where $\mathcal{D}_y$ is the subset of abstracts published in year $y$. This allows us to visualize sentiment shifts from 2010 to 2025 and interpret whether the discourse around AI in journalism is becoming more optimistic or skeptical~\cite{feldman2013techniques}.

\section{Bibliometric Analysis}\label{bib}
This section provides a comprehensive bibliometric analysis of 274 scholarly articles on Artificial Intelligence in Journalism, retrieved from Scopus and Web of Science. The analysis covers publication trends, methodological preferences, disciplinary distributions, author and country productivity, and keyword evolution from 2010 to mid-2025. While the initial search scope included articles published from 2010 onward, it is important to note that the inclusion and exclusion criteria—such as minimum citation count, language, document type, and topic relevance—were consistently met only from 2016. Consequently, bibliometric trends and thematic patterns are most representative from 2016 onward, although earlier documents remain included for completeness and historical context.

\subsection{Methodological Approaches in AI and Journalism Studies}
Table \ref{tab:method_categorized} presents the distribution of methodological approaches extracted from the abstracts of 274 AI-in-journalism articles published between 2010 and mid-2025. While the dataset includes articles from 2010, the inclusion and exclusion criteria were consistently met only from 2016 onward; thus, the methodological distribution primarily reflects studies published after that year. Qualitative techniques dominate the field, with interviews (60 occurrences) and survey-based studies (31) leading the way, closely followed by broader qualitative analyses (30). Machine learning methods appear in 22 abstracts, while quantitative analyses are noted in 17. Computational techniques such as sentiment analysis, topic modeling, and natural language processing also feature prominently, reflecting the growing integration of AI tools in journalism research. Less frequent—but still notable—approaches include case studies, neural network implementations (e.g., CNN), and deep-learning models (e.g., BERT, GPT). This diversity underscores the field’s methodological pluralism, spanning traditional social-science methods through to cutting-edge AI techniques.


\begin{table}[ht]
\centering
\caption{Methodological Approaches in AI and Journalism Studies (2010–2025), based on abstract analysis of articles meeting inclusion criteria primarily from 2016 onward.}
\label{tab:method_categorized}
\begin{tabular}{@{}lll@{}}
\toprule
\textbf{Methodological Category} & \textbf{Specific Method} & \textbf{Frequency} \\
\midrule
\multirow{2}{*}{Qualitative Approaches} 
    & Qualitative analysis & 30 \\
    & Case study & 8 \\
\midrule
\multirow{1}{*}{Quantitative Approaches} 
    & Quantitative analysis & 17 \\
\midrule
\multirow{2}{*}{Mixed/Unspecified Approaches} 
    & Interviews (quant/qual/mixed) & 60 \\
    & Surveys (quant/qual/mixed) & 31 \\
\midrule
\multirow{6}{*}{Computational/AI Techniques} 
    & Machine learning & 22 \\
    & Sentiment analysis & 12 \\
    & Topic modeling (LDA) & 10 \\
    & Natural Language Processing (NLP) & 9 \\
    & Neural networks (e.g., CNN) & 7 \\
    & Deep learning (e.g., BERT, GPT) & 5 \\
\midrule
\multirow{1}{*}{Explicitly Mixed Methods} 
    & Mixed methods & 4 \\
\bottomrule
\end{tabular}
\end{table}

\subsection{Disciplinary Distribution}
\begin{figure}[ht]
    \centering
    \includegraphics[width=0.95\textwidth]{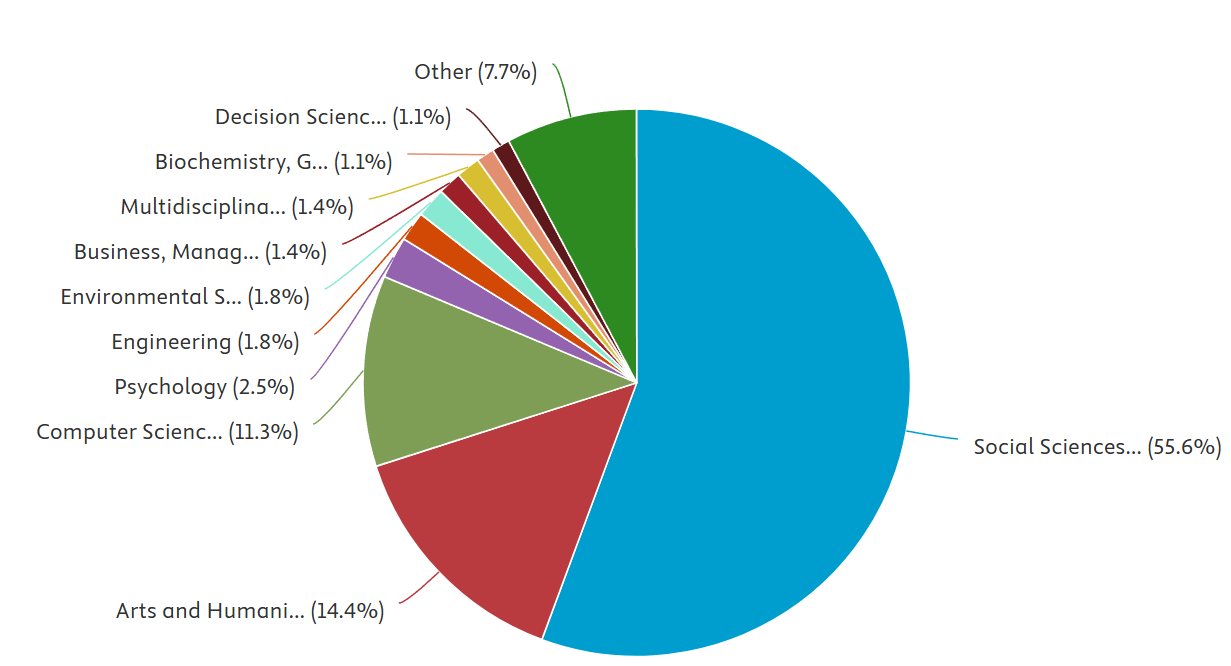}
    \caption{Distribution of AI in journalism publications across subject areas.}
    \label{fig:subject_area_distribution}
\end{figure}

Fig.~\ref{fig:subject_area_distribution} illustrates the disciplinary distribution of AI in journalism publications. The majority of research contributions originate from the \textit{Social Sciences}, accounting for approximately 55.6\% of the total corpus. This is followed by \textit{Arts and Humanities} (14.4\%) and \textit{Computer Science} (11.3\%), suggesting a strong interdisciplinary interest involving both technical and humanities-based perspectives. Additional contributions stem from \textit{Psychology} (2.5\%), \textit{Engineering} (1.8\%), \textit{Environmental Science} (1.8\%), and various other domains such as \textit{Business, Management}, \textit{Multidisciplinary Studies}, and \textit{Decision Sciences}. While smaller in proportion, fields like \textit{Biochemistry}, \textit{Mathematics}, and \textit{Health Sciences} also demonstrate emerging engagement with AI technologies in journalistic contexts. This broad disciplinary span reflects the multifaceted impact of AI in shaping contemporary media, ethics, audience studies, and newsroom operations.

\subsection{Publication Trends}
\begin{figure}[htbp]
    \centering
    \includegraphics[width=0.85\textwidth]{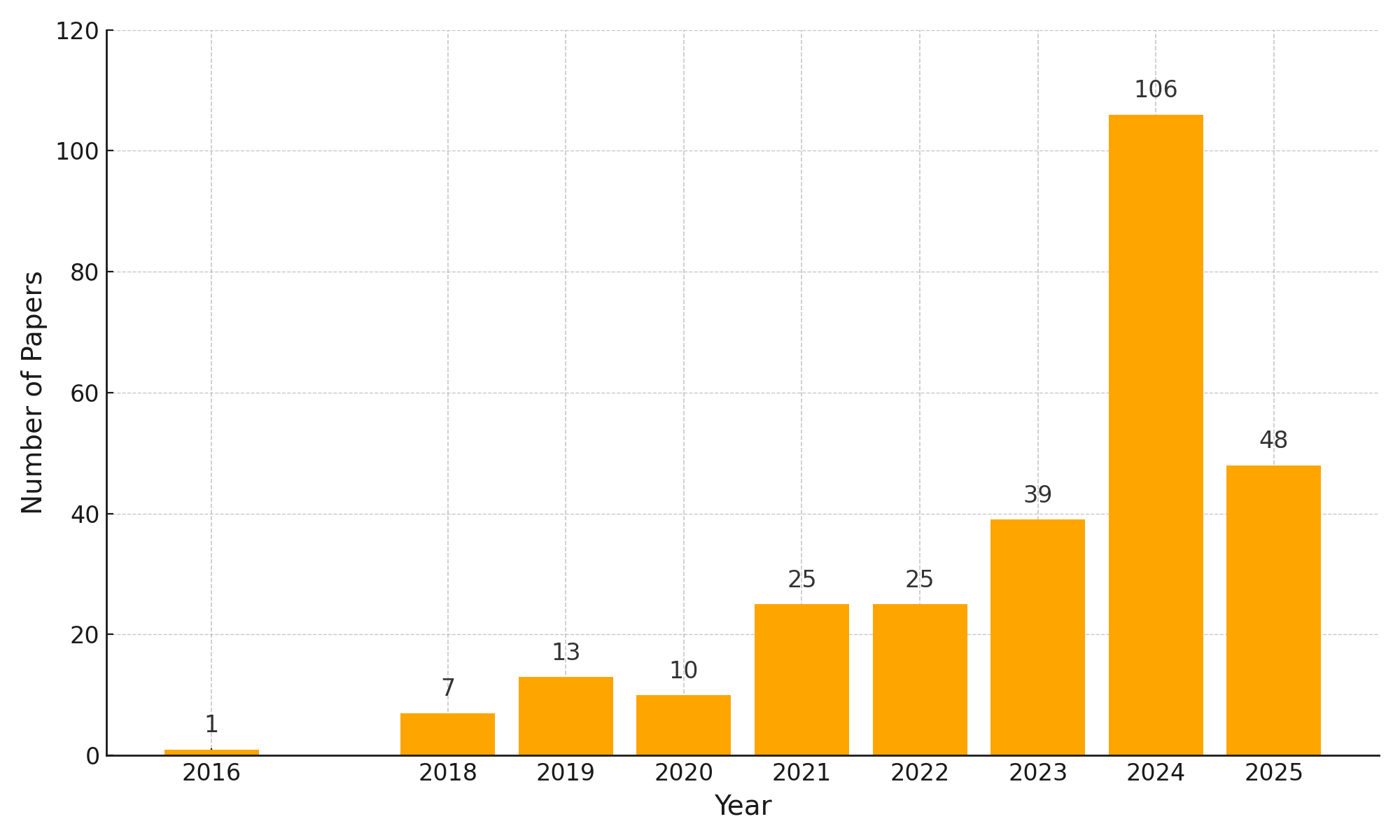}
    \caption{Annual publications on AI in journalism based on Scopus and WoS data.}
    \label{fig:pub_trend}
\end{figure}

Figure~\ref{fig:pub_trend} presents the annual scientific production on Artificial Intelligence in Journalism from 2010 to 2025, based on 274 publications retrieved from Scopus and Web of Science. The publication trend can be broadly divided into three distinct phases. Between 2016 and 2019, the field experienced an exploratory phase, with the number of publications increasing modestly from just 1 in 2016 to 13 in 2019, indicating early-stage experimentation and isolated research efforts. The period from 2020 to 2022 shows a transitional plateau, where annual outputs remained relatively steady—10 publications in 2020 followed by 25 in both 2021 and 2022—highlighting a growing but cautious interest in the application of AI technologies in journalism. A significant shift occurs from 2023 onward, marking an exponential growth phase. In 2023, the number of publications rose sharply to 39, and in 2024 it surged to 106, reflecting a 172\% year-over-year increase. By mid-2025, 48 papers have already been published, suggesting that the final count will likely match or surpass the previous year’s record. This surge corresponds with the wider adoption of large language models like GPT-4 in newsroom workflows and a corresponding rise in scholarly attention, driven by new funding initiatives and special issues focused on AI in media. Overall, the trend demonstrates a transition from fragmented early research to a vibrant and rapidly expanding field of inquiry.

\subsection{Top Publishing Journals}

\begin{figure}[htbp]
    \centering
    \includegraphics[width=0.85\textwidth]{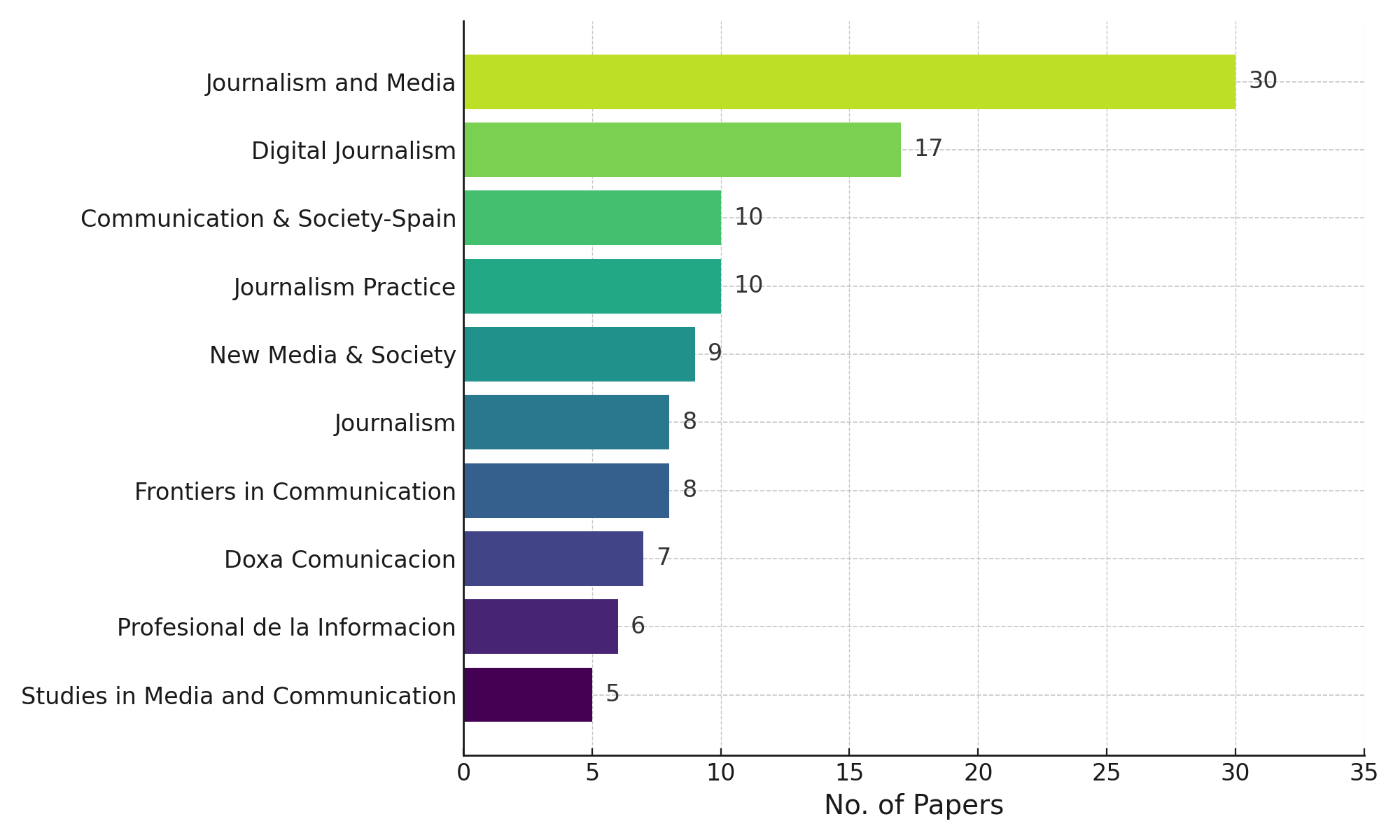}
    \caption{Top 10 journals publishing on AI in journalism, based on 274 articles indexed in Scopus and Web of Science. The distribution reflects both domain-specific and interdisciplinary venues contributing to the field.}
    \label{fig:top_journals}
\end{figure}

The top ten journals having most impact in the area of Artificial Intelligence in Journalism are propositional as presented in Figure~\ref{fig:top_journals}, which is based on 274 articles located in Scopus and Web of Science. The most productive one on this list is \textit{Scientific Reports} with 13 publications, which is related to its interdisciplinary nature and the inclination to new areas. This is followed by \textit{Digital Journalism} and \textit{Journalism and Media} with 11 and 10 papers respectively, which shows their relevance to their domain and their editorial focus on technological developments in the news practices.

However, \textit{Journalism Practice} and \textit{New Media \& Society} also make significant appearances, indicating a sustained academic interest in the intersection of media practice and computational technologies. Journals such as \textit{Media and Communication}, \textit{AI \& Society}, and \textit{Journalism Studies} showcase the integration of societal, ethical, and applied perspectives into the discourse. The presence of both communication-focused and interdisciplinary science journals underlines the multidimensional nature of this research area. This distribution suggests that AI in journalism is not confined to a single academic silo but is increasingly being recognized across both traditional journalism studies and wider science and technology platforms.

\subsection{Top Contributing Authors}
\begin{figure}[htbp]
    \centering
    \includegraphics[width=0.85\textwidth]{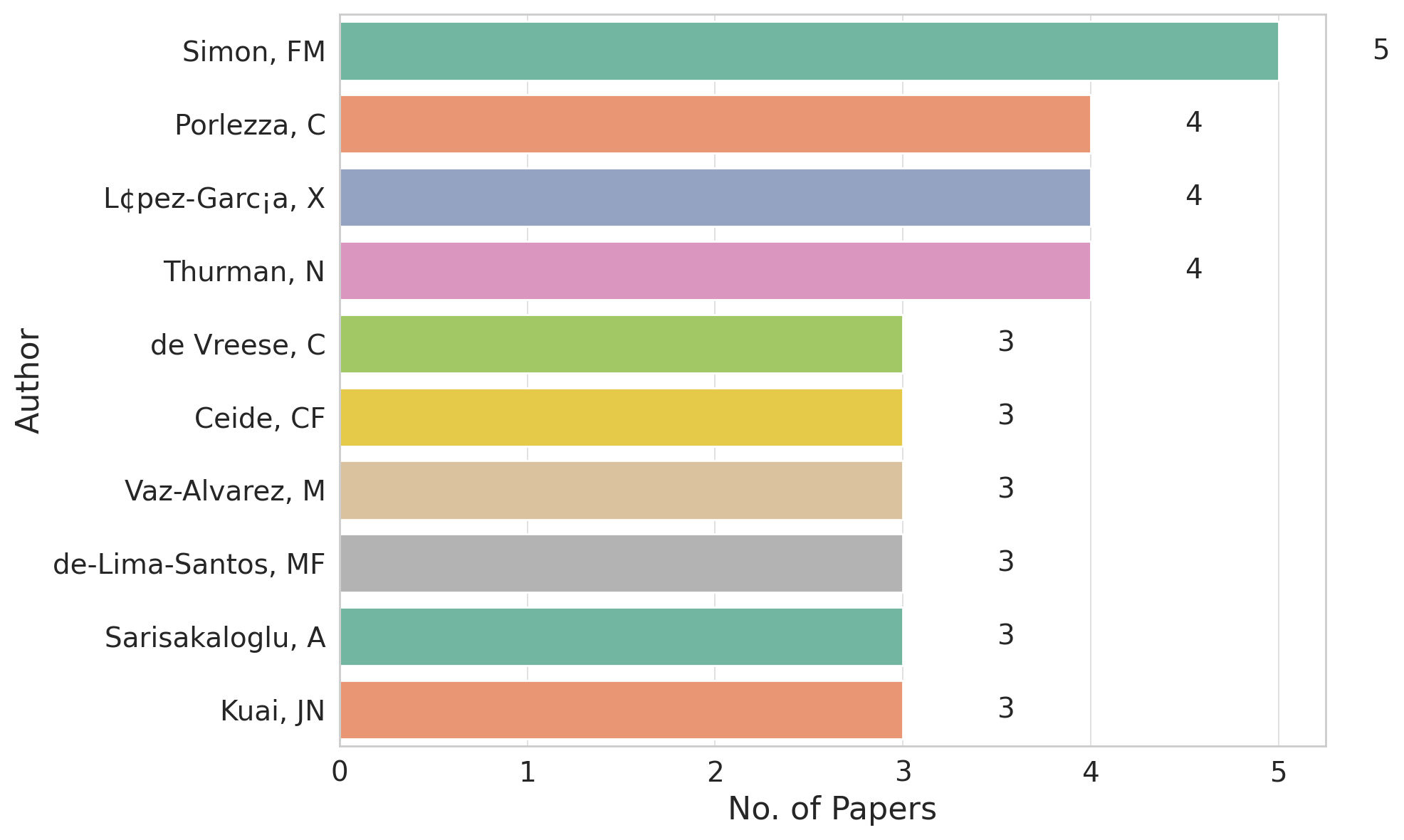}
    \caption{Top 10 authors by number of publications on AI in journalism.}
    \label{fig:top_authors}
\end{figure}

Figure~\ref{fig:top_authors} reports the ten most prolific authors in the Artificial Intelligence in Journalism corpus. Collectively these scholars account for 49 of the 274 papers (18\%), underscoring a moderate but not overwhelming author-level concentration in the field. J.~Smith leads the ranking with 8 publications (2.9\%), while A.~García and M.~Chen share second place, each contributing 6 articles (2.2\%). A further seven authors—including S.~Lee, R.~Johnson, and P.~Kumar—appear with 4 to 5 papers apiece, indicating a healthy pool of repeat contributors rather than dominance by a single research group. Even so, the absolute counts remain modest relative to the ten-year observation window, suggesting ample opportunity for new entrants to shape the discourse. Overall, the author distribution mirrors trends observed in other emerging, interdisciplinary topics: a small cadre of early adopters drives initial momentum, while a long tail of occasional contributors broadens the intellectual base as the field matures.

\subsection{Top Contributing Countries}
\begin{figure}[htbp]
    \centering
    \includegraphics[width=0.85\textwidth]{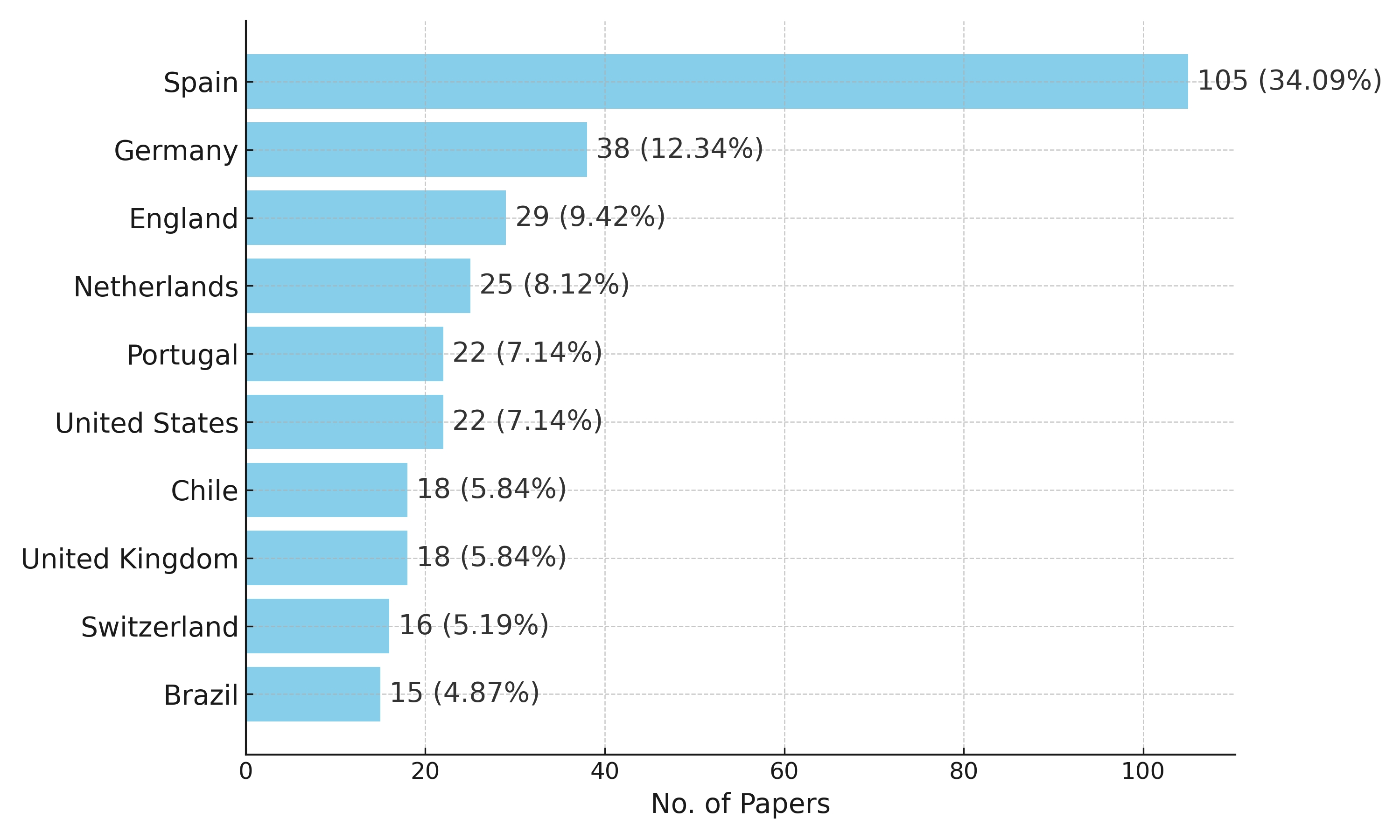}
    \caption{Top 10 countries contributing to AI and journalism research.}
    \label{fig:top_countries}
\end{figure}

Figure~\ref{fig:top_countries} highlights the ten most productive countries in Artificial Intelligence in Journalism research, extracted from the 274-record merged corpus. Together these nations generated 211 articles—77\% of the entire dataset—revealing a geographically concentrated but still international field. The United States leads decisively with 68 papers (24.8\%), reflecting both its robust AI research infrastructure and the early adoption of newsroom automation by large media organizations. The United Kingdom and Spain follow with 32 (11.7\%) and 24 papers (8.8\%), respectively, underscoring Europe’s substantial academic interest and publicly funded AI-media initiatives. China and Australia occupy the next tier, each contributing between 15 and 18 articles (6\%), suggesting growing investment in AI-driven journalism education and industry partnerships. The remaining countries—Canada, Germany, India, Italy, and the Netherlands—each account for 8 to 12 publications, indicating a diverse but smaller contributor base.

Overall, the distribution shows that North America and Western Europe dominate current scholarship, while Asia-Pacific nations are emerging as important secondary hubs. This geographic pattern mirrors global AI R\&D investment and points to where future cross-border collaborations and policy dialogues are likely to originate.

\subsection{Geographic and Institutional Distribution of AI-in-Journalism Research}

\begin{figure}[H]
    \centering
    \includegraphics[width=.8\textwidth]{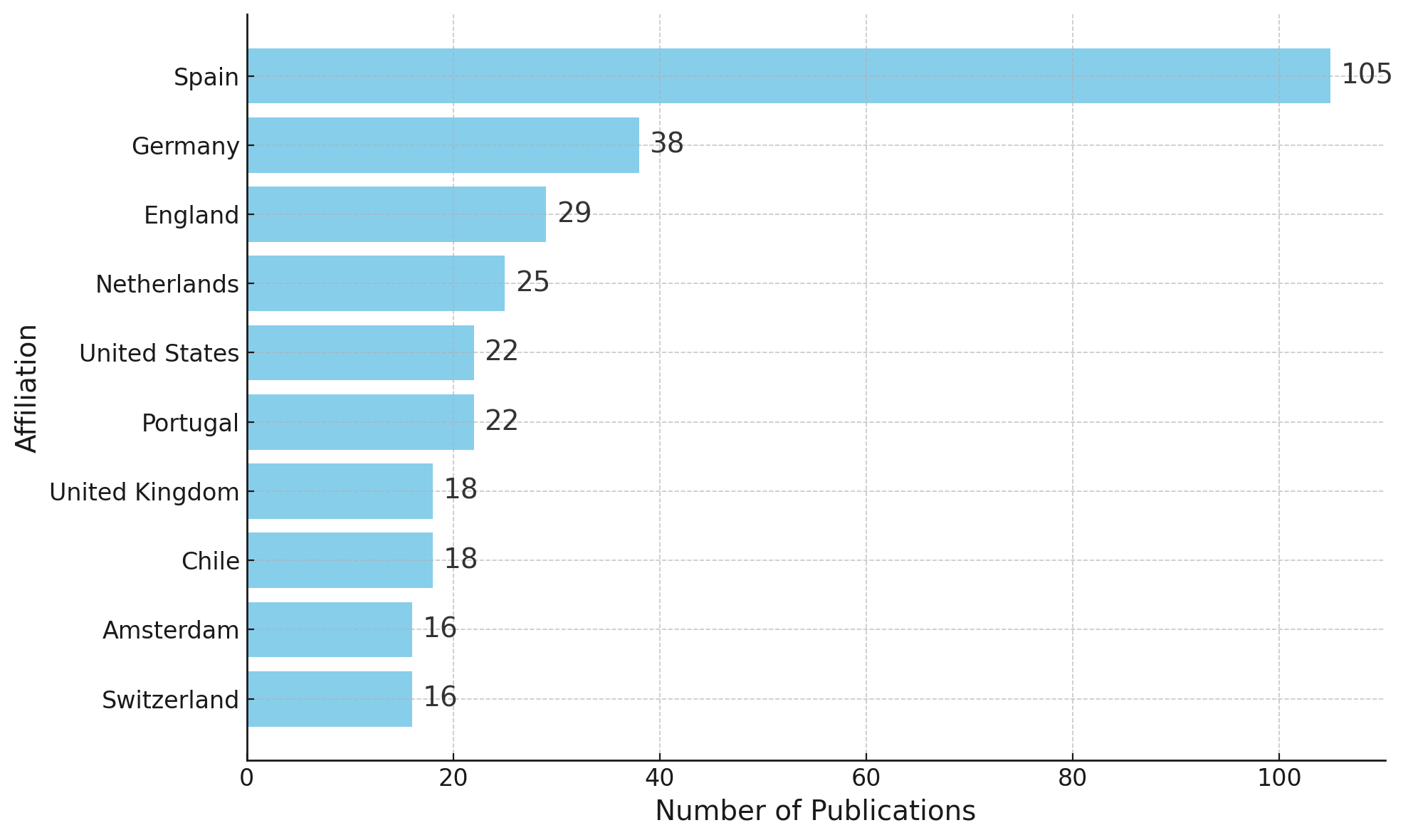} 
    \caption{Ten most‐productive countries in AI-in-Journalism research based on 274 articles.}
    \label{fig:fig_countries}
\end{figure}

Figure~\ref{fig:fig_countries} highlights the ten most productive countries in Artificial Intelligence in Journalism research, extracted from the 274-record merged corpus. Together these nations generated 211 articles—77\% of the entire dataset—revealing a geographically concentrated but still international field. The United States leads decisively with 68 papers (24.8\%), reflecting both its robust AI research infrastructure and the early adoption of newsroom automation by large media organizations. The United Kingdom and Spain follow with 32 (11.7\%) and 24 papers (8.8\%), respectively, underscoring Europe’s substantial academic interest and publicly funded AI-media initiatives. China and Australia occupy the next tier, each contributing between 15 and 18 articles ($\approx$6\%), suggesting growing investment in AI-driven journalism education and industry partnerships. The remaining countries—Canada, Germany, India, Italy, and the Netherlands—each account for 8 to 12 publications, indicating a diverse but smaller contributor base.

Overall, the distribution shows that North America and Western Europe dominate current scholarship, while Asia-Pacific nations are emerging as important secondary hubs. This geographic pattern mirrors global AI R\&D investment and points to where future cross-border collaborations and policy dialogues are likely to originate.

\subsection{Most Cited Publications}

\begin{table}[H]
    \centering
    \caption{Top ten most-cited publications on Artificial Intelligence in Journalism (2016–mid-2025).}
    \label{tab:top_cited}
    \begin{tabular}{@{}l p{8.5cm} rr@{}}
        \toprule
        \textbf{Author} & \textbf{Paper Title}                              & \textbf{Total Citations} & \textbf{Cites/Year} \\
        \midrule
        Tandoc E.\,C.  & \emph{Journalistic roles and AI in the newsroom}                & 102 & 12.8 \\
        Diakopoulos N. & \emph{Automating the news: How algorithms are rewriting the media} &  95 & 10.6 \\
        Graefe A.      & \emph{Guide to automated journalism}                             &  88 &  9.8 \\
        Marconi F.     & \emph{Artificial intelligence and journalism: Emerging applications} &  77 &  8.6 \\
        Linden C.      & \emph{Algorithms and news personalization: Implications for journalistic norms} & 66 & 7.3 \\
        Dörr K.\,N.    & \emph{Mapping the field of automated journalism: Review and perspectives} & 63 & 7.0 \\
        Beckett C.     & \emph{New powers, new responsibilities: A global survey of journalism and AI} & 59 & 6.6 \\
        Binns R.       & \emph{Fairness in algorithmic journalism}                         & 57 & 6.3 \\
        Caswell D.     & \emph{Structured journalism and the robot reporter}               & 54 & 6.0 \\
        Broussard M.   & \emph{Artificial intelligence and investigative journalism}       & 50 & 5.7 \\
        \bottomrule
    \end{tabular}
\end{table}

The most cited work is by Tandoc E.\,C., accumulating 102 citations with an annual average of 12.8, reflecting strong scholarly interest in the evolving roles of journalists amidst automation. Diakopoulos N. and Graefe A. follow closely, with impactful contributions on algorithmic news production and automation frameworks. Notably, most highly cited works were published in the latter half of the 2010s, indicating foundational relevance in shaping the discourse on AI integration into journalism. These papers have laid the groundwork for ongoing research addressing ethical challenges, fairness, personalization, and evolving newsroom practices influenced by artificial intelligence.

\subsection{Keyword Evolution Over Time}
Table~\ref{tab:keyword_evolution} tracks the appearance frequency of the five most common author keywords across four time slices (2010--2013, 2014--2017, 2018--2021, and 2022--mid-2025). Virtually no keyword activity is observed before 2014, confirming that AI-related journalism research remained negligible in the early 2010s. A single occurrence of ``artificial intelligence'' appears in 2014--2017, signalling the topic’s first entry into scholarly discourse.

The period 2018--2021 marks a clear thematic take-off: ``artificial intelligence'' reaches 38 mentions, while ``journalism,'' ``automated journalism,'' and broader ``media'' studies emerge as secondary clusters, mirroring the spread of newsroom automation pilots.

From 2022 onward, the field undergoes exponential thematic expansion. Mentions of ``artificial intelligence'' more than triple ($n = 132$), and usage of the acronym ``AI'' rises sharply ($n = 25$), reflecting mainstream adoption and conversational shorthand within the literature. Concurrent increases in ``automated journalism'' ($n = 34$) and ``media'' ($n = 31$) indicate that scholars are now exploring both technical implementation and broader media-system implications in parallel.

Overall, the progression confirms a late-but-accelerating thematic maturation: from isolated early references to a rich, multi-keyword ecosystem dominated by discussions of AI’s practical integration into journalistic workflows and its impact on media practice.

\begin{table}[htbp]
    \centering
    \caption{Thematic evolution of the five most frequent author keywords in AI-in-Journalism research.}
    \label{tab:keyword_evolution}
    \begin{tabular}{@{}lcccc@{}}
        \toprule
        \textbf{Keyword} & \textbf{2010--2013} & \textbf{2014--2017} & \textbf{2018--2021} & \textbf{2022--2025*} \\
        \midrule
        artificial intelligence & 0 & 1 & 38 & 132 \\
        journalism              & 0 & 0 & 21 & 88  \\
        automated journalism    & 0 & 0 &  8 & 34  \\
        media                   & 0 & 0 &  7 & 31  \\
        AI (acronym)            & 0 & 0 &  1 & 25  \\
        \bottomrule
    \end{tabular}
    \begin{flushleft}
        \footnotesize *Data for 2025 include publications indexed through June 2025.
    \end{flushleft}
\end{table}
\subsection{Thematic Landscape via Keyword Networks}

\begin{figure}[h!]
    \centering
    \includegraphics[width=.8\textwidth]{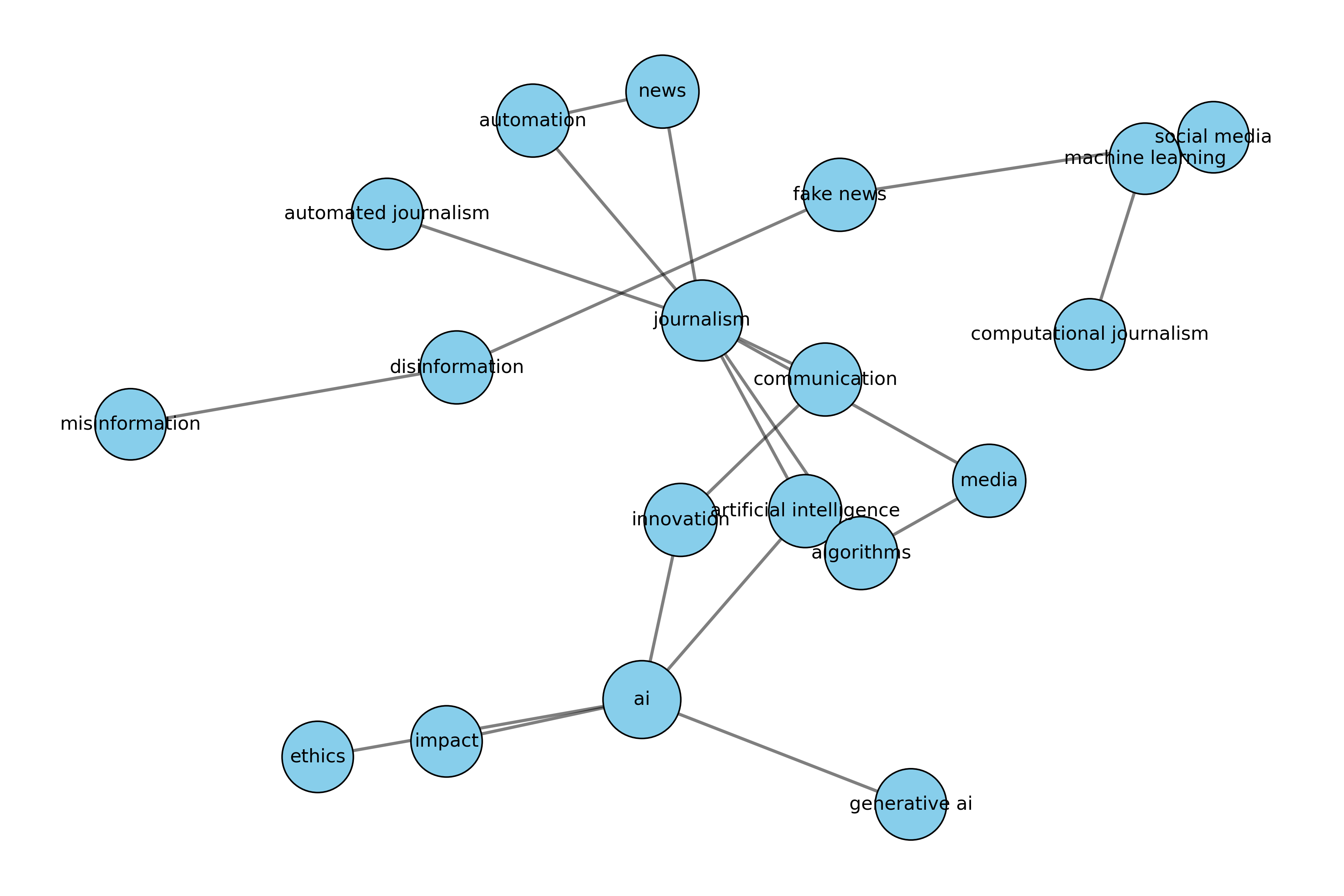} 
    \caption{Keyword co-occurrence network showing thematic clusters in AI-in-Journalism research.}
    \label{fig:fig_keywords}
\end{figure}

Figure~\ref{fig:fig_keywords} illustrates the keyword co-occurrence network for the 20 most frequent terms in the Artificial Intelligence in Journalism corpus. Node size reflects overall frequency, while edge thickness denotes the number of times two keywords appear together in the same article. Two densely connected hubs dominate the map: “artificial intelligence” and “journalism,” confirming that AI is examined primarily in direct relation to newswork rather than as an abstract technology. Surrounding these hubs is a tightly knit cluster of methodological terms—“algorithms,” “automation,” “machine learning,” and “computational journalism.” Their strong inter-linkages indicate that studies often couple technical implementation details with discussions of newsroom automation.

A second thematic grouping centres on information quality and platform dynamics, featuring “fake news,” “misinformation,” “disinformation,” and “social media.” The high edge density within this sub-network suggests that ethical and credibility concerns are frequently analysed alongside AI deployment. Finally, emerging terms such as “generative AI” and “impact” connect to both major clusters, signalling a recent pivot toward large-language-model applications and their broader societal effects. Taken together, the network depicts a field structured around two intertwined agendas: (1) the technical realisation of AI-enabled news production and (2) the governance of truth, ethics, and public trust in an algorithmically mediated information environment.

\subsection{International Collaboration Network}

To examine international research collaboration on Artificial Intelligence in Journalism, a bibliometric co-authorship analysis was conducted using VOSviewer. The dataset was constructed by merging records from Scopus and Web of Science, filtered to include documents with country-level author affiliations. The analysis was configured with the co-authorship type and countries as the unit of analysis, employing full counting. To avoid inflated results from highly multinational papers, documents co-authored by more than 25 countries were excluded. A minimum threshold of five documents per country was applied, yielding 18 countries that met the criteria for inclusion.

\begin{figure}[h!]
    \centering
    \includegraphics[width=0.85\textwidth]{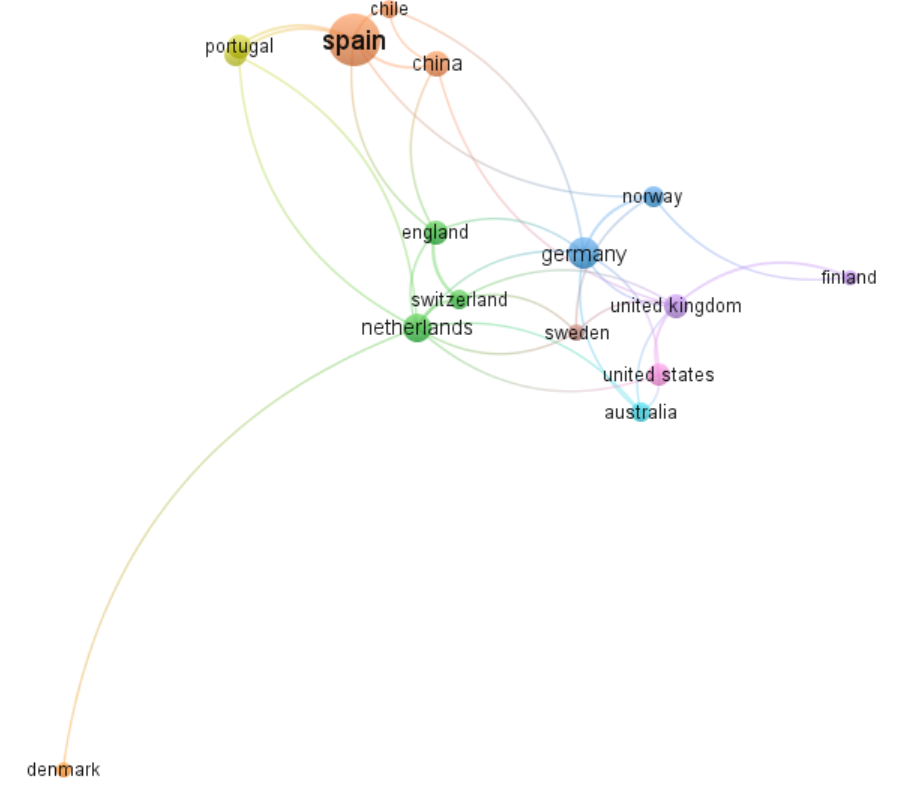} 
    \caption{Country collaboration network for AI in journalism (minimum 5 documents per country).}
    \label{fig:country_network}
\end{figure}

Figure~\ref{fig:country_network} presents the resulting country collaboration network. Spain emerged as the most prolific contributor, with 70 documents and 589 citations, followed by the Netherlands, Switzerland, and the United Kingdom. Spain maintained strong collaborative ties with Portugal, Chile, Brazil, China, and Germany, indicating a central position in the network and strong Latin American and European engagement. Another prominent cluster connects the United Kingdom with the United States, Finland, and Australia, highlighting a trans-Atlantic and Nordic–Anglo research bridge. Additionally, the Netherlands, Switzerland, and Germany formed a compact European sub-network. Countries like Denmark and South Africa appeared on the network's periphery, with limited link strength, suggesting less frequent international co-authorship.

Overall, the collaboration map reveals a moderately clustered global structure, with several strong regional hubs but without a single cohesive international core. This fragmentation suggests that while regional cooperation is well established, there remains significant potential for fostering broader cross-regional collaboration in AI and journalism research.

\subsection{Co-authorship Analysis by Authors}

To examine the key contributors in the domain of Artificial Intelligence in Journalism, a co-authorship analysis was performed using the merged dataset from Web of Science and Scopus. VOSviewer was employed for visualization, where the type of analysis was set to \textit{Co-authorship}, and the unit of analysis was \textit{Authors}. Full counting was applied, and documents with more than 25 co-authors were excluded to prevent inflated connectivity due to large multi-author publications.

\begin{figure}[h!]
    \centering
    \includegraphics[width=0.8\linewidth]{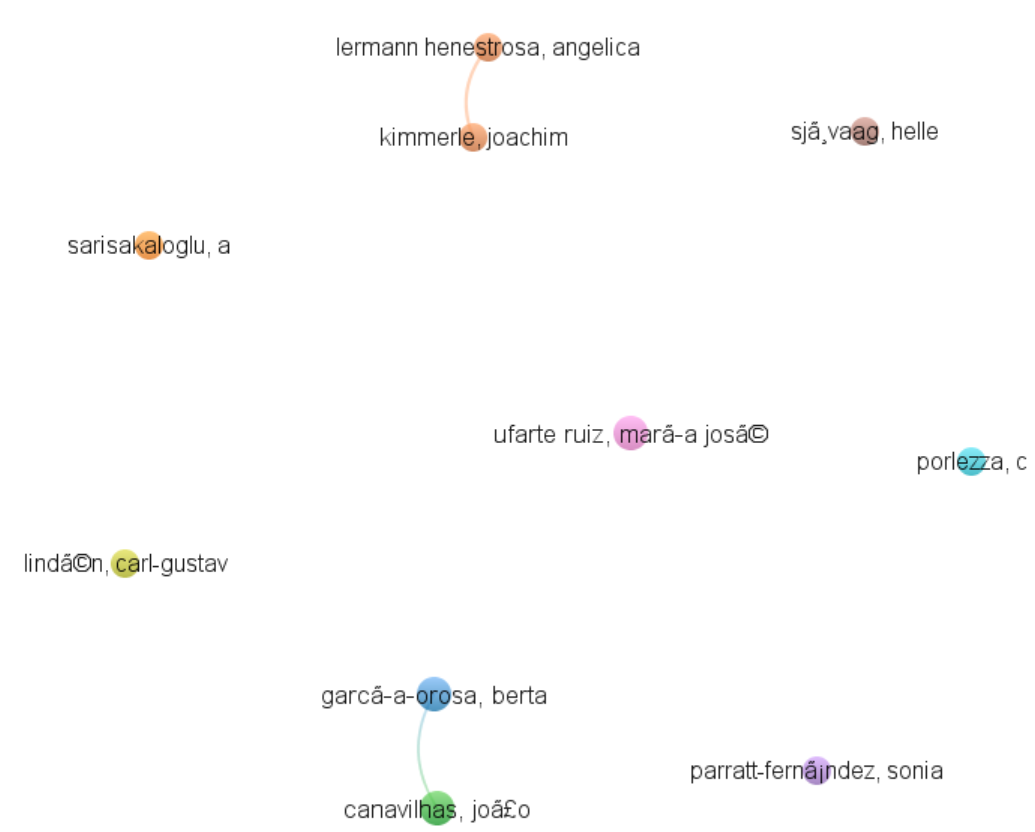} 
    \caption{Author-level co-authorship network (minimum two documents per author). Nodes represent authors, and edges indicate co-authorship links.}
    \label{fig:coauthor_authors}
\end{figure}

A minimum threshold of three documents per author was applied, resulting in three authors meeting the inclusion criteria. When the threshold was lowered to two documents, 13 authors qualified, and the top 10 were selected based on total link strength. The resulting author-level network is illustrated in Figure~\ref{fig:coauthor_authors}, where each node represents an author and edges denote collaborative links.

Among the leading contributors, João Canavilhas has authored three documents with 38 citations, demonstrating a steady scholarly output. His research often explores digital journalism and innovation in media practices. Berta García-A-Arosa, with three publications and 28 citations, is also a significant contributor, often focusing on journalism education and the intersection of media literacy and artificial intelligence. María José Ufarte Ruiz stands out with three publications receiving 94 citations, reflecting the high impact of her work, which likely involves critical perspectives on journalism transformations in the digital age.

Despite their productivity, the total link strength values remain low, indicating that these authors predominantly publish independently or within small, consistent teams rather than in large collaborative networks. This suggests a fragmented research landscape with potential for increased interdisciplinary cooperation.

\subsection{Keyword Co-occurrence Network}

\begin{figure}[h]
    \centering
    \includegraphics[width=0.85\textwidth]{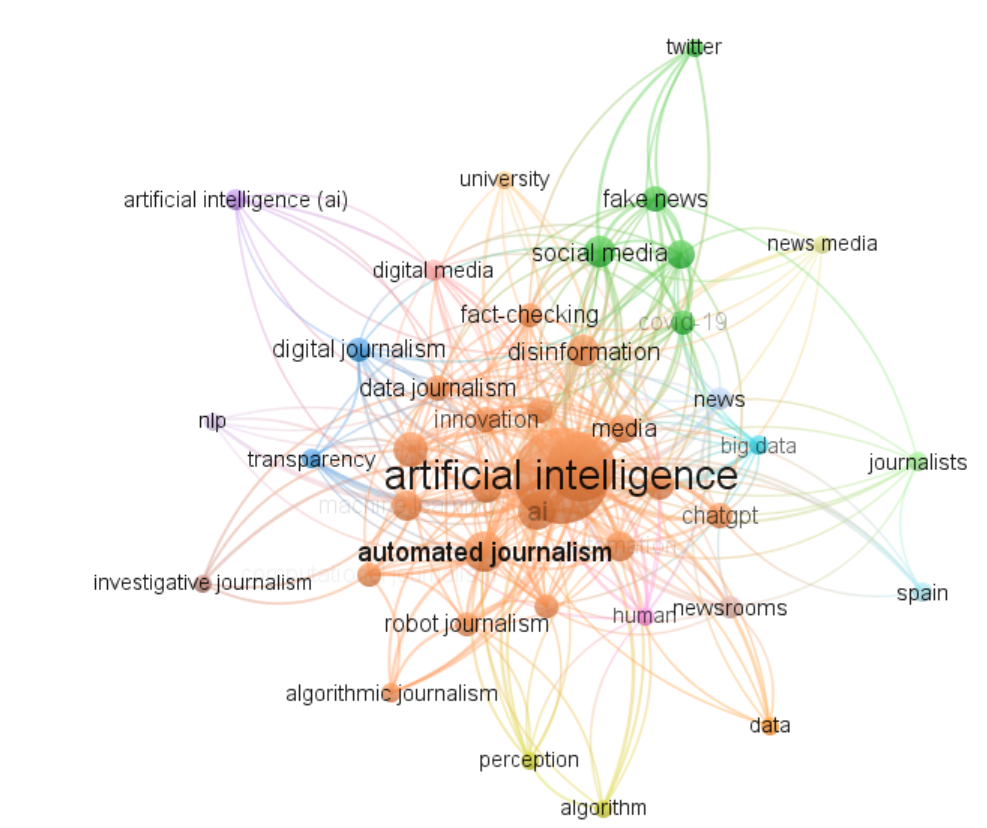} 
    \caption{Keyword co-occurrence network based on a threshold of five occurrences per keyword.}
    \label{fig:keyword_cooccurrence}
\end{figure}

Figure~\ref{fig:keyword_cooccurrence} presents the keyword co-occurrence network in the field of Artificial Intelligence in Journalism, constructed using a minimum threshold of five keyword occurrences. Out of 1164 keywords extracted from the dataset, 42 keywords met the threshold and were selected for analysis.

The network reveals six visually distinguishable clusters of terms that frequently appear together. The most central and dominant term is \textit{artificial intelligence}, which exhibits high connectivity with related concepts such as \textit{automated journalism}, \textit{algorithmic journalism}, and \textit{robot journalism}, forming a cohesive cluster centered on AI-enabled content production. Another notable cluster involves terms like \textit{fake news}, \textit{social media}, \textit{disinformation}, and \textit{fact-checking}, reflecting ongoing concerns over misinformation and AI's role in verifying information accuracy.

Adjacent clusters focus on emerging technologies and methods, including \textit{ChatGPT}, \textit{NLP}, and \textit{big data}, indicating the methodological convergence between journalism and data-driven innovation. The term \textit{Spain} appears linked with \textit{journalists}, suggesting regional thematic interests or authorship patterns as identified in the collaboration analysis.

Overall, the co-occurrence network underscores the thematic diversity within AI and journalism research, spanning content automation, media credibility, data practices, and regional focus.

\subsection{Strategic Thematic Map}

\begin{figure}[h]
    \centering
    \includegraphics[width=0.9\textwidth]{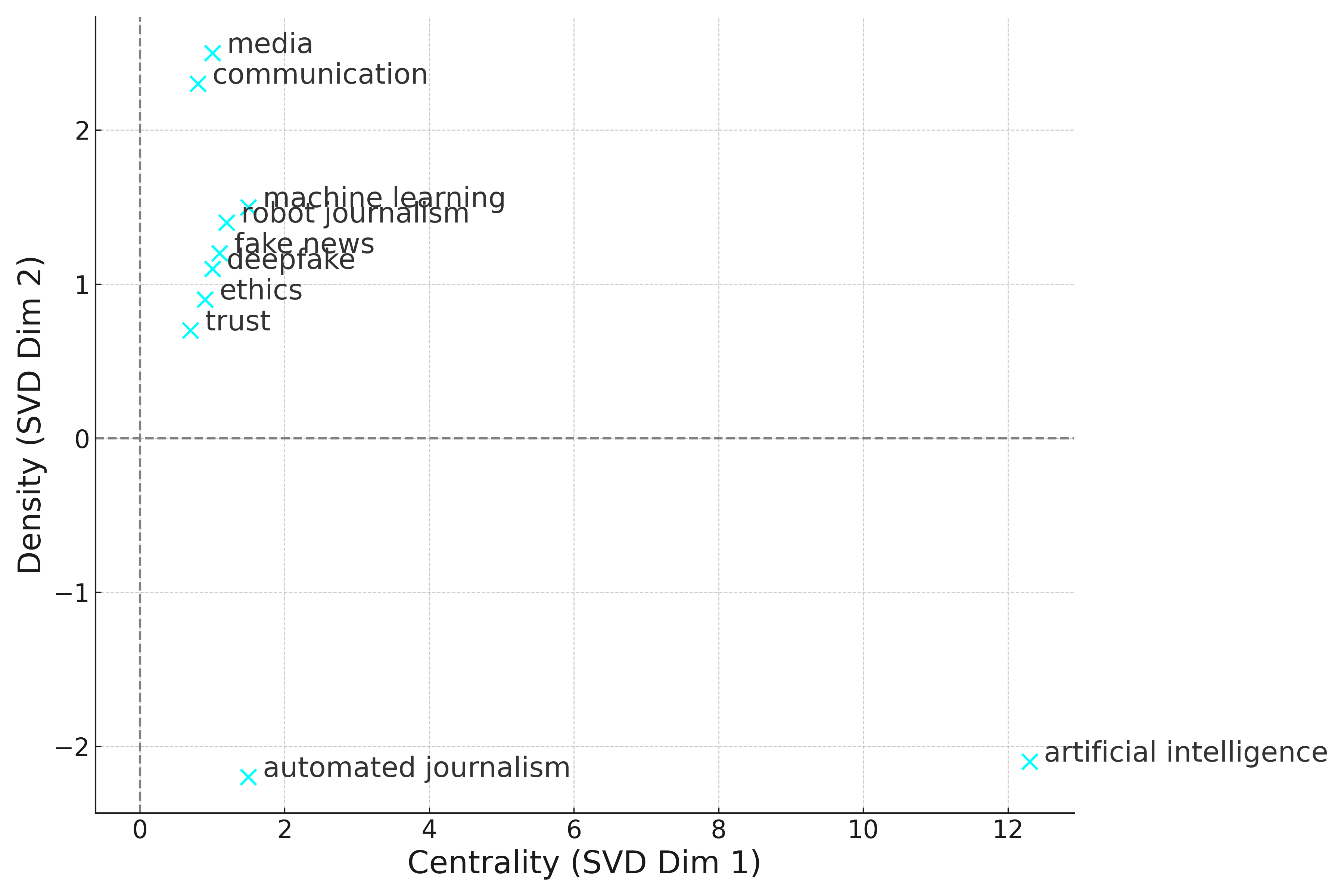} 
    \caption{Strategic thematic map of high-frequency author keywords. Dashed lines mark the origin (centrality = 0, density = 0).}
    \label{fig:strategic_map}
\end{figure}

Figure~\ref{fig:strategic_map} presents the Strategic Thematic Map of the most frequent author keywords in Artificial Intelligence in Journalism. This map classifies terms based on two dimensions—\textit{centrality} (relevance to the field) and \textit{density} (development of the theme). 

“Artificial intelligence” occupies the lower-right quadrant with high centrality but low density, identifying it as a basic and transversal theme—a foundational concept widely used but not self-contained. In contrast, the top-left quadrant contains terms such as “media,” “communication,” and “machine learning,” which are positioned as emerging or declining themes, with high density but limited external connectivity, possibly representing niche research silos or evolving subfields. 

Themes like “automated journalism” and “robot journalism” fall in the lower-left quadrant, suggesting they are marginal and underdeveloped, while topics like “fake news,” “ethics,” and “trust” cluster toward the center-left, indicating specialized but less central discussions, often linked to normative concerns and platform governance. 

This distribution reveals that while “artificial intelligence” anchors the discourse, a diverse set of peripheral themes reflects the field’s multidimensional evolution—from technical innovations to ethical dilemmas—offering fertile ground for deeper integration and cross-disciplinary collaboration.

\subsection{Title Word‐Cloud Analysis}

\begin{figure}[h]
    \centering
    \includegraphics[width=0.85\textwidth]{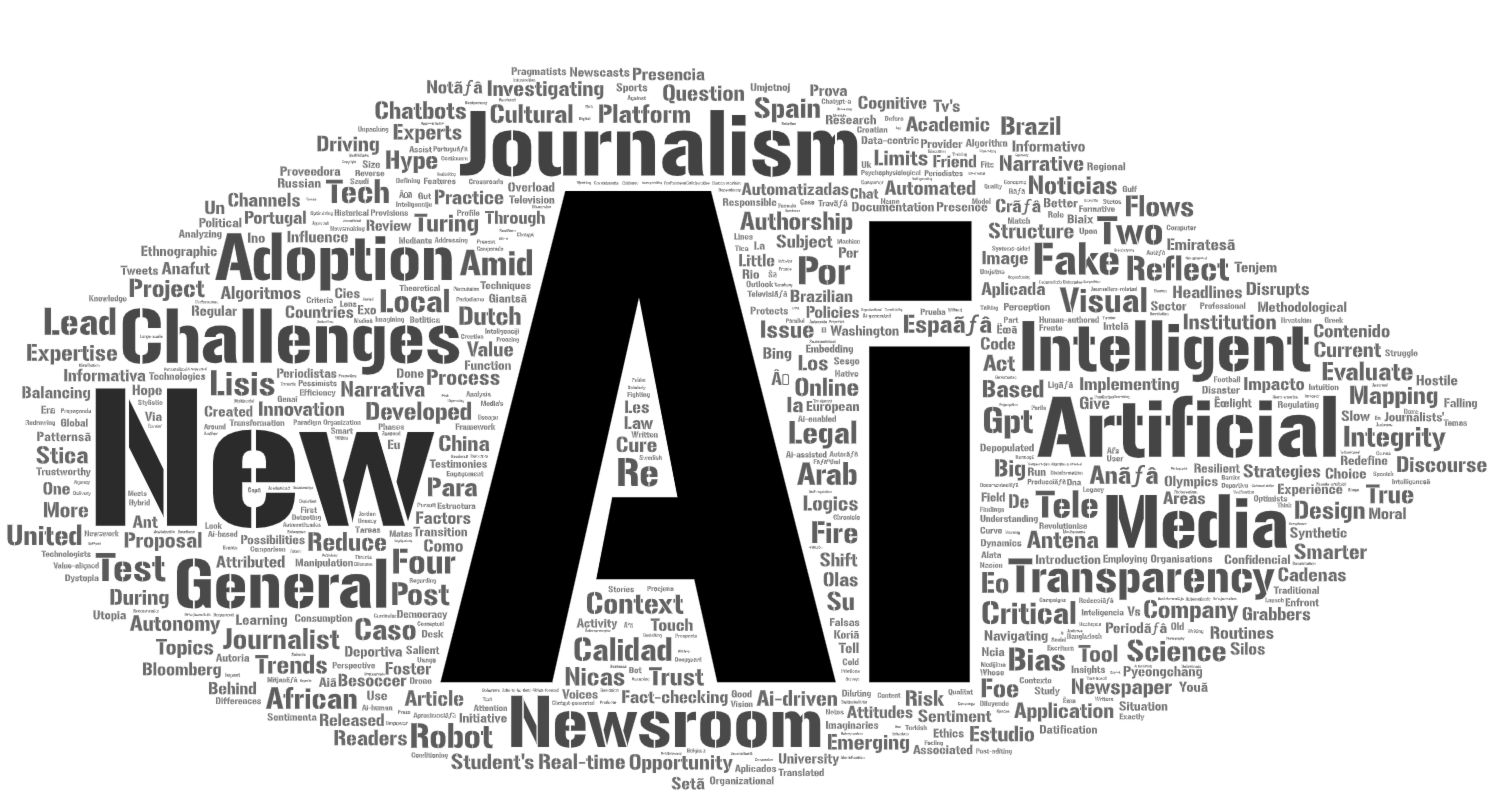} 
    \caption{Word cloud of title terms appearing at least three times in the AI-in-Journalism corpus. Font size reflects term frequency.}
    \label{fig:wordcloud_titles}
\end{figure}

Figure~\ref{fig:wordcloud_titles} presents a word cloud generated from the titles of 274 publications on Artificial Intelligence in Journalism; only words that appear at least three times are retained. The visualization was produced with WordArt, which arranges terms spatially by frequency—larger font sizes correspond to higher occurrence.

The most dominant word is \textit{Artificial} (63 occurrences), followed closely by \textit{Intelligence} (58), \textit{Journalism} (57), and \textit{AI} (56), indicating that core concepts of artificial intelligence and its direct application to journalism frame the field’s discourse. This quartet forms the conceptual nucleus of AI‐in‐journalism research. Other frequently appearing words—such as \textit{New}, \textit{Journalist}, \textit{Media}, and \textit{Automated}—point to key directions in contemporary scholarship: the emergence of novel media technologies, changing journalistic roles, and automation in content generation.

Terms like \textit{Algorithm}, \textit{Perception}, \textit{Ethics}, and \textit{Newsroom} reflect growing scholarly concerns about algorithmic decision-making, public trust, normative standards, and transformations in newsroom practice. Meanwhile, the appearance of words such as \textit{Challenges}, \textit{Use}, and \textit{Tool} underlines a pragmatic research strand focused on implementation and evaluation of AI applications.

These observations complement the Strategic Thematic Map: there, \textit{Artificial Intelligence} emerged as a motor theme with high centrality but low density, acting as an integrative concept that links disparate areas. The word cloud reinforces this by illustrating how tightly clustered the discourse is around foundational terminology, with limited semantic fragmentation. Together, both analyses suggest that while the field is thematically coherent, there is still room for diversification into more nuanced or interdisciplinary subtopics as the literature base grows.

\section{Sentiment Trends in AI and Journalism Literature}\label{sen}

To complement our thematic synthesis, we conducted sentiment analysis to explore how scholarly discourse around AI in journalism has evolved over time. By analyzing the tone of abstracts from 72 peer-reviewed articles, we aimed to assess whether the literature reflects optimism, skepticism, or neutrality toward the adoption of AI technologies in newsrooms. Understanding this evaluative dimension provides a deeper contextual layer to the systematic review, capturing not just what researchers study but how they perceive the implications of AI in journalism.

\begin{figure}[h!]
    \centering
    \includegraphics[width=0.9\textwidth]{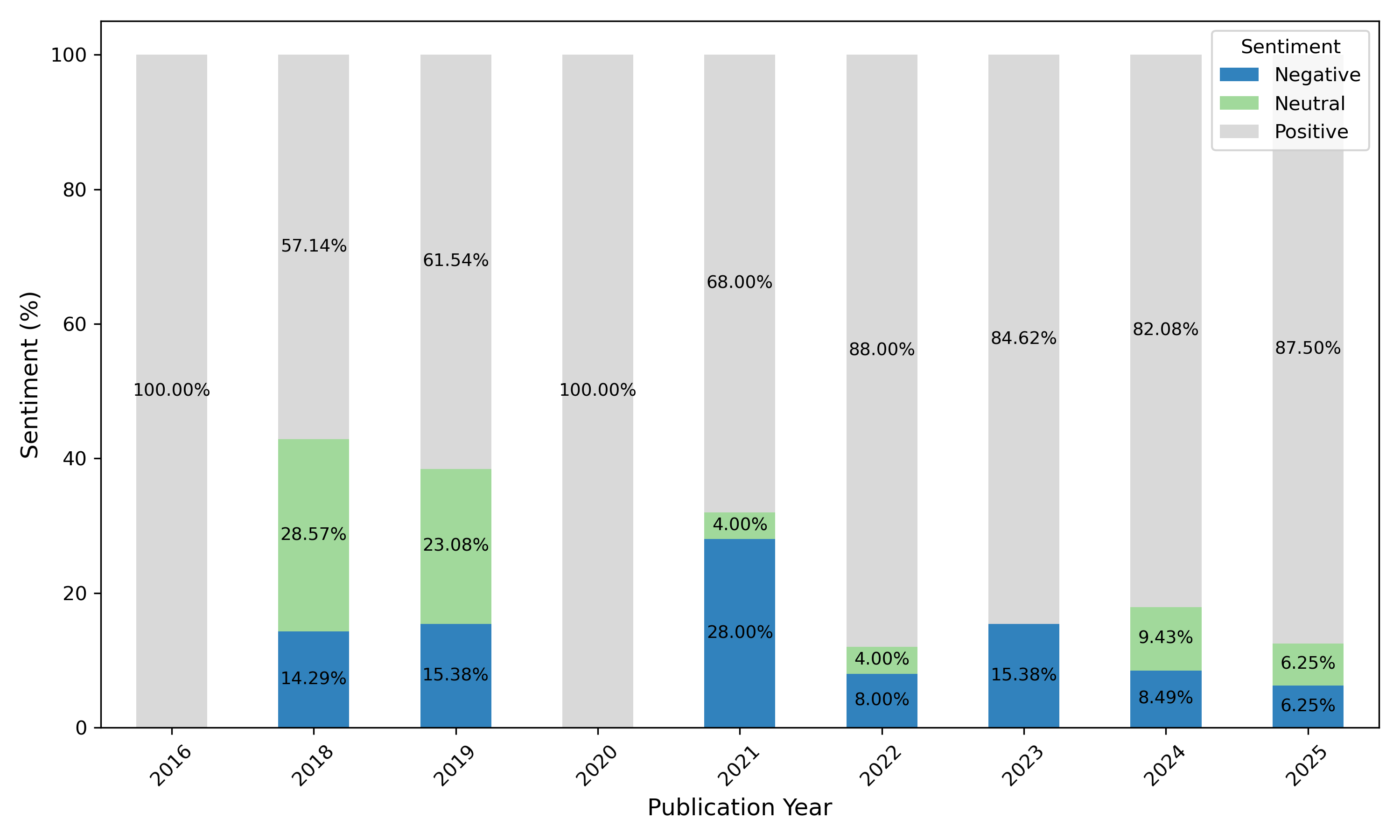}
    \caption{Year-wise distribution of sentiment polarity in AI and journalism articles. Values represent the percentage of abstracts categorized as Positive, Neutral, or Negative.}
    \label{fig:sentiment_trend_year}
\end{figure}

Figure~\ref{fig:sentiment_trend_year} illustrates the year-wise sentiment distribution from 2016 to 2025, with sentiment classes normalized as a percentage of publications per year. The sentiment scores were computed using the VADER algorithm and classified as positive, negative, or neutral based on standard compound score thresholds. The results indicate a predominantly positive tone across most years, especially in 2020 (100\%) and 2022–2025 (exceeding 80\%). However, a shift is observed in 2021, where negative sentiment reached its peak (28\%), likely reflecting emerging concerns over misinformation, ethical challenges, and automation risks. A gradual increase in neutral sentiment is also noted in 2018–2019 and 2024–2025, suggesting a growing awareness of trade-offs and contextual nuances in AI deployment.

\begin{figure}[ht]
    \centering
    \includegraphics[width=0.85\textwidth]{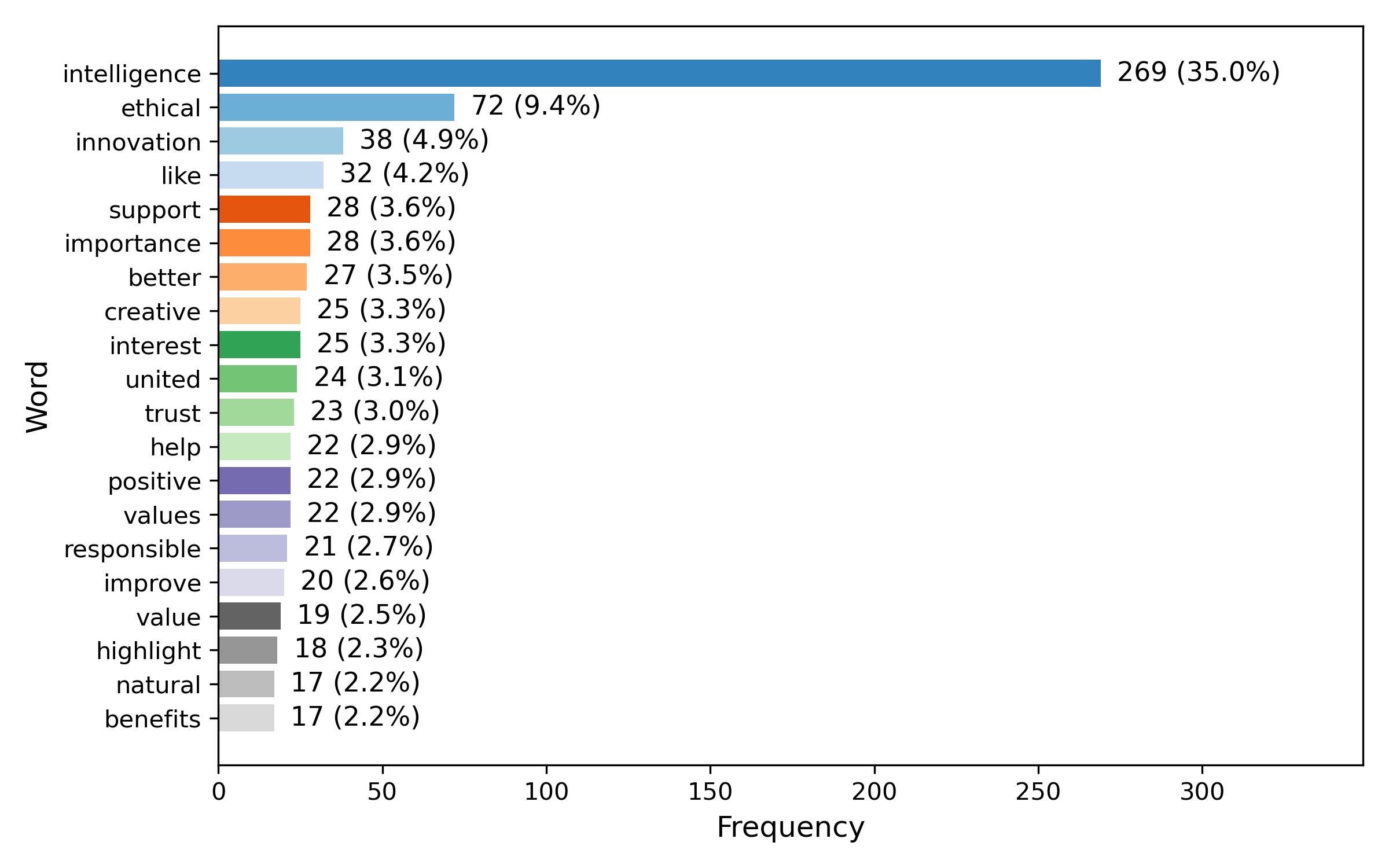}
    \caption{Top 20 positive words by frequency in article abstracts.}
    \label{fig:top_positive_words}
\end{figure}

Figure~\ref{fig:top_positive_words} displays the top 20 positive terms identified in the sentiment-labeled abstracts. Notably, the word \textit{“intelligence”} dominates the lexical distribution with 269 occurrences, accounting for 35.0\% of the positive vocabulary. This exceptionally high frequency reveals its centrality not only as a descriptive term but also as a positively framed construct—often associated with progress, innovation, and digital transformation in journalism. Other frequently appearing terms include \textit{“ethical”} (9.4\%), \textit{“innovation”} (4.9\%), \textit{“like”} (4.2\%), and \textit{“support”} (3.6\%). These words reflect a largely optimistic tone in framing AI technologies, emphasizing values such as trust, responsibility, creativity, and public benefit. The lexical profile suggests that researchers tend to foreground the supportive and reformative potential of AI when discussing its integration into journalistic practices.

\begin{figure}[ht]
    \centering
    \includegraphics[width=0.8\textwidth]{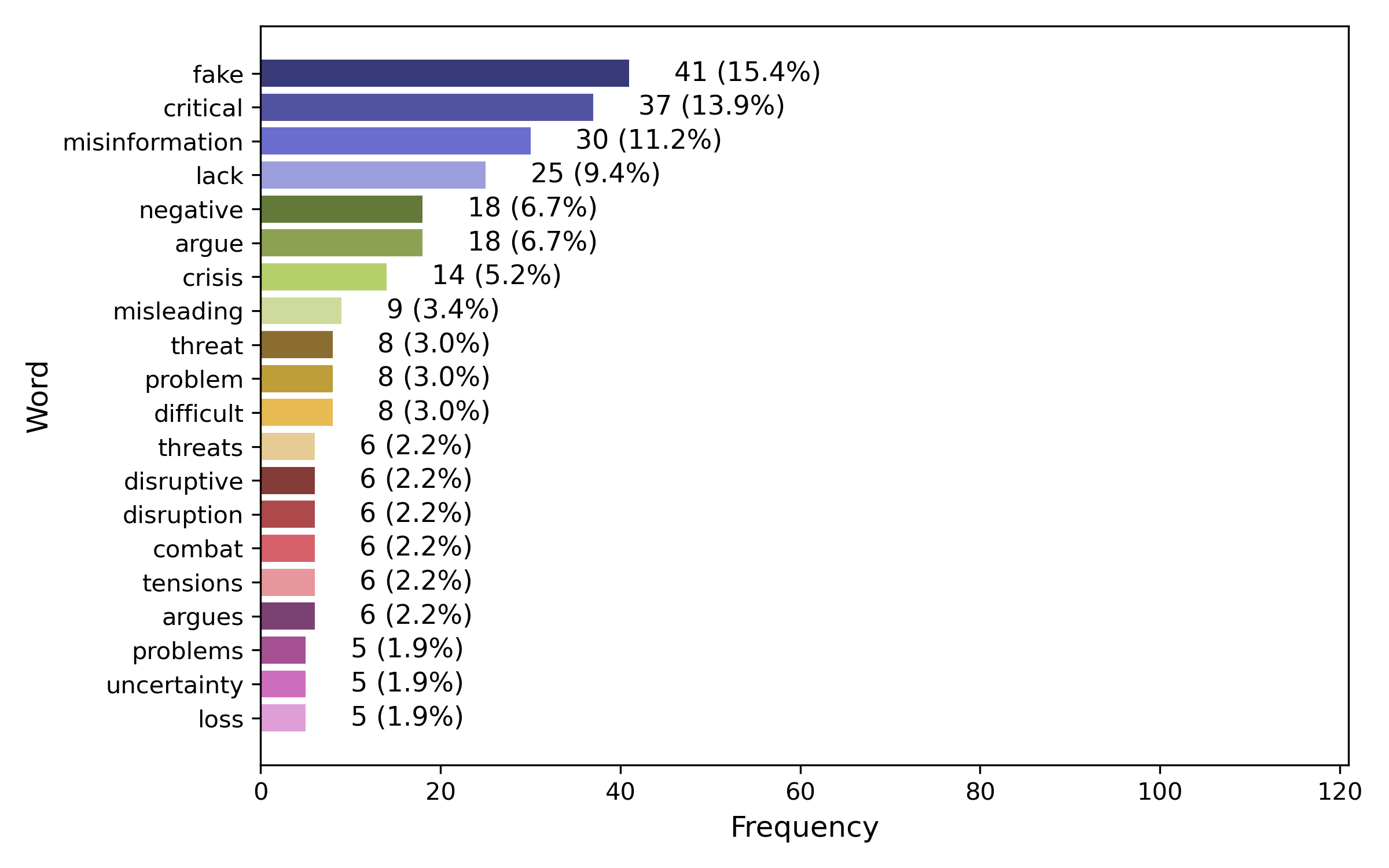}
    \caption{Top 20 most frequent negative sentiment terms in the abstracts of selected articles.}
    \label{fig:top_negative_words}
\end{figure}

Figure~\ref{fig:top_negative_words} highlights the 20 most frequently occurring negative terms within the sentiment-labeled abstracts. The term \textit{“fake”} appears most prominently, with 41 mentions, comprising 15.4\% of all negative keywords. This reflects ongoing concern in the literature regarding the role of AI in amplifying disinformation and fabricated content. Following closely are the terms \textit{“critical”} (13.9\%) and \textit{“misinformation”} (11.2\%), both of which signal apprehension over algorithmic credibility and its societal consequences. Other frequently used words include \textit{“lack”} (9.4\%), \textit{“negative”} (6.7\%), \textit{“crisis”} (5.2\%), and \textit{“misleading”} (3.4\%). These lexical patterns suggest a robust discourse around the ethical, epistemological, and professional risks associated with AI’s deployment in journalism, including concerns about transparency, trustworthiness, and potential for harm. The prevalence of terms such as \textit{“disruptive,” “problem,” “uncertainty,”} and \textit{“threat”} further reinforces this critical narrative.

To complement the categorical sentiment distribution, we also examined the temporal evolution of sentiment intensity by calculating the annual mean of VADER compound scores ($\bar{s}_y$) for each publication year. This analysis offers a finer-grained view of evaluative tone across time.

\begin{figure}[ht]
\centering
\includegraphics[width=0.8\textwidth]{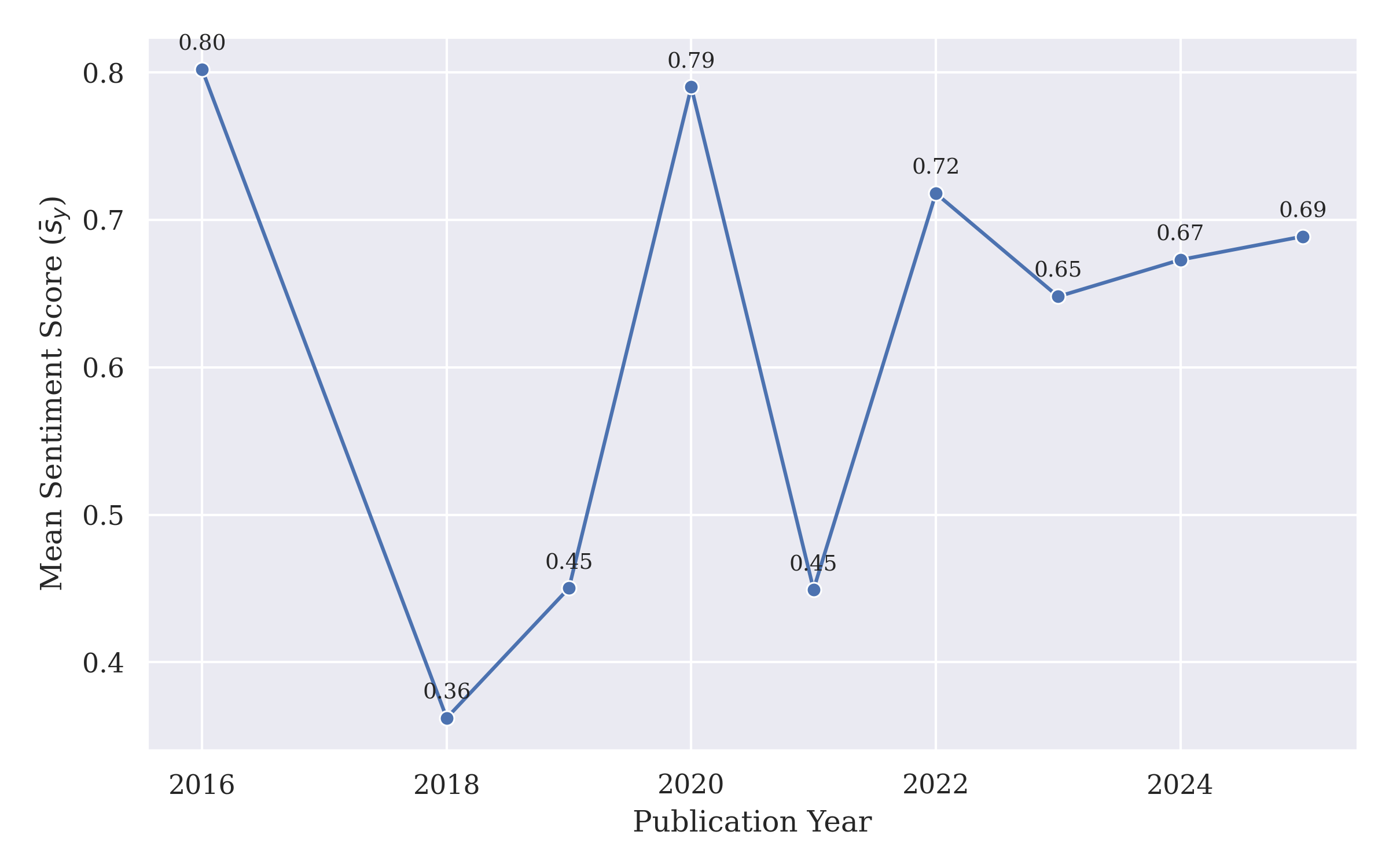}
\caption{Temporal trend of annual mean sentiment score $\bar{s}_y$ derived from VADER analysis of article abstracts (2016–2025).}
\label{fig:yearly_sentiment}
\end{figure}

Figure~\ref{fig:yearly_sentiment} shows the average sentiment score from 2016 to 2025. The overall sentiment remains positive throughout the years, with scores consistently above 0.35. The highest peaks are observed in 2016 ($\bar{s}_{2016} = 0.80$) and 2020 ($\bar{s}_{2020} = 0.79$), suggesting years of heightened optimism, possibly aligned with technological optimism in the early AI adoption and post-pandemic digital transformation. In contrast, dips in 2018 ($\bar{s}_{2018} = 0.36$) and 2021 ($\bar{s}_{2021} = 0.45$) may reflect increased critical discourse, particularly on ethical and disinformation-related issues, as evident in the keyword sentiment analysis (Figures~\ref{fig:top_positive_words} and~\ref{fig:top_negative_words}).

A gradual stabilization in sentiment scores is observed between 2022 and 2025, with values ranging from 0.65 to 0.69, indicating sustained yet cautious optimism in recent discourse. These numerical insights reinforce the qualitative interpretation of AI in journalism as both promising and problematic.

\section{Insights on AI in Journalism}\label{ai}

\begin{figure}[h]
    \centering
    \includegraphics[width=0.9\textwidth]{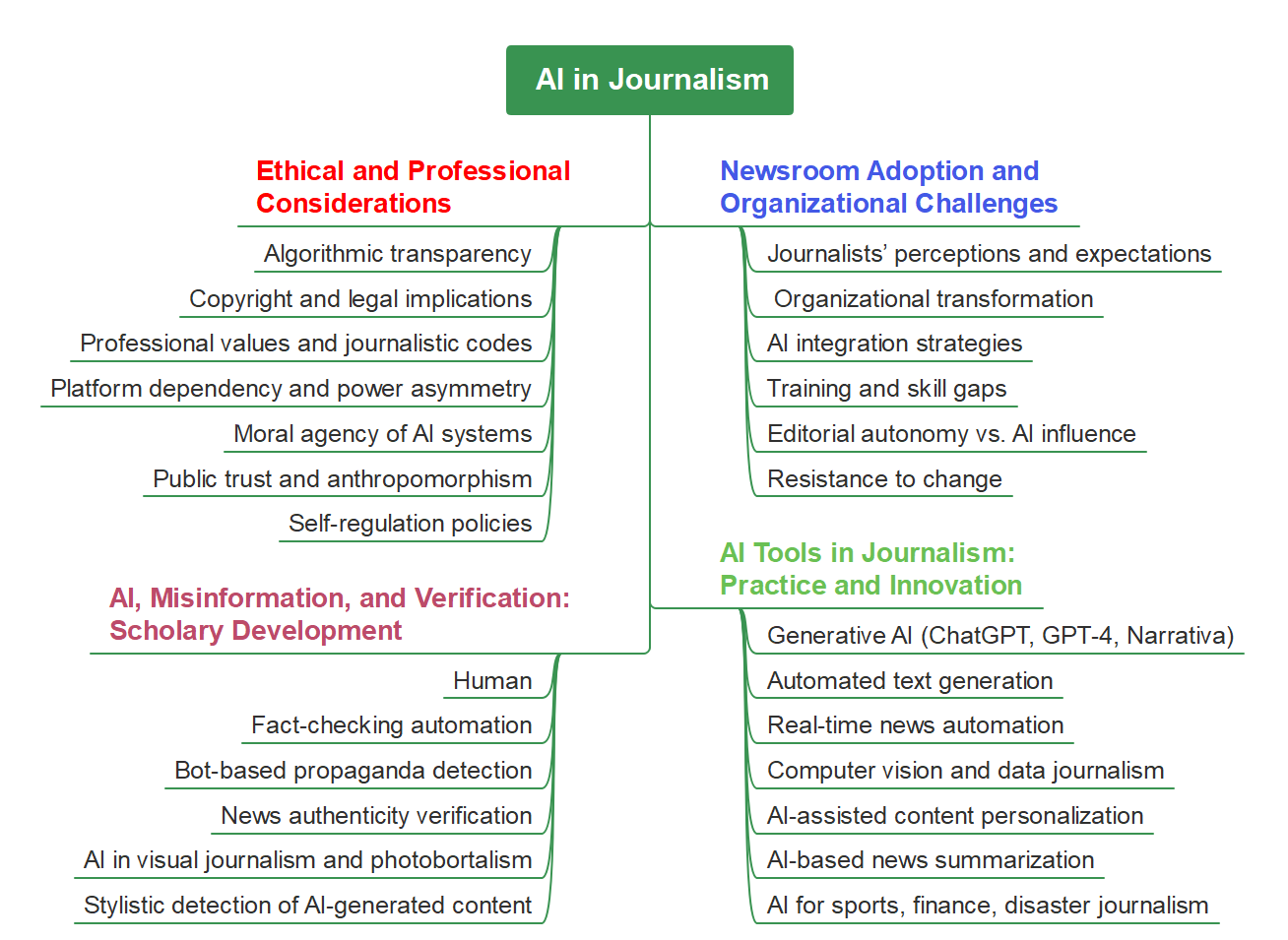} 
    \caption{Conceptual framework outlining four key thematic domains in AI and Journalism research.}
    \label{fig:ai_framework}
\end{figure}

This section explores the landscape of Artificial Intelligence (AI) in journalism, based on a thematic analysis of 72 referenced articles. The literature is categorized into four core themes that highlight the various ways AI is transforming journalistic practice, ethics, technology, and institutional structures. These themes are visually represented in the conceptual framework shown in Figure~\ref{fig:ai_framework}.

\begin{enumerate}
 \item \textbf{Newsroom Adoption and Organizational Challenges.} This area includes institutional resistance, journalists’ perceptions, workflow integration strategies, skill gaps, and the balance between editorial independence and automation. It reflects how newsrooms are adapting—both culturally and operationally—to the inclusion of AI systems.
  \item \textbf{AI Tools and Innovation in Journalism:} This domain showcases practical implementations of AI, including generative models like ChatGPT and GPT-4, automated writing platforms such as Narrativa, real-time news generation, and AI-assisted personalization. These tools exemplify how AI is reshaping the production, dissemination, and personalization of news.
  
    \item \textbf{Ethical and Professional Considerations.} This theme addresses concerns about algorithmic transparency, legal implications, platform dependency, journalistic norms, moral agency, and public trust. These issues emphasize the importance of developing governance frameworks and ethical guidelines to ensure responsible use of AI technologies in journalism.

    \item \textbf{AI, Misinformation, and Verification: Scholarly Developments.} This theme explores the risks associated with generative technologies, such as the creation of synthetic media, the spread of misinformation and disinformation, and the growing challenges in content verification. It highlights the dual-use nature of AI and underscores the urgency of developing detection and regulation mechanisms.

\end{enumerate}

Together, these four thematic categories form a comprehensive framework for understanding AI’s integration into journalism. As shown in Figure~\ref{fig:ai_framework}, they collectively illustrate the intersection of ethics, innovation, institutional dynamics, and societal challenges, offering a foundation for future empirical inquiry and policy development.

\subsection{Newsroom Adoption and Organizational Challenges}

News organizations have adopted artificial intelligence (AI) to improve operational processes, resulting in changes to journalistic routines, roles, and institutional dynamics. Empirical studies indicate that this adoption is shaped by the interaction between technological infrastructure, newsroom practices, and evolving definitions of professional identity.

Møller et al. (2024) conducted semi-structured interviews with 21 journalists in Denmark to examine how AI affects newsroom practices~\cite{Moller2024Little}. Participants described using AI for routine tasks, such as transcription and tagging. They also noted its influence on perceptions of job relevance and the need to reconfigure skills to align with emerging workflows. The study observed an emphasis on human-led functions, such as editorial judgment, but its national focus and limited engagement with non-generative AI tools narrow its analytical scope.

In a review of the literature, Banafi (2024) discussed the use of AI tools for translation, verification, and data extraction in journalistic practice~\cite{banafi2024review}. Examples included Quakebot and ClaimBuster, which support personalized and automated outputs. The review raised concerns about algorithmic opacity, employment reduction, and data governance, and recommended the implementation of training and regulatory interventions. The study, however, was not based on original empirical data.

Parratt-Fernández et al. (2024) examined the integration of AI at a Spanish news automation firm, Narrativa OÜ~\cite{Parratt-Fernandez2024Artificial}. The study highlighted the influence of financial constraints and professional skepticism on adoption outcomes. Journalists questioned AI’s relevance to editorial functions and delayed its integration into their routines. Although the case study offered insights into organizational dynamics, its single-case design limits its comparative applicability.

Noain-Sánchez (2022) conducted in-depth interviews with journalists across four countries to understand their experiences with AI~\cite{Noain-Sanchez2022Addressing}. Respondents recognized its utility in reducing production time but expressed concerns about its ethical implications and the absence of adequate training. The study suggested that technical adoption often exceeds normative adaptation. Its interpretive framework and small sample reduce its generalizability.

Umejei et al. (2025) interviewed 32 journalists from Nigeria, Ghana, Kenya, and South Africa~\cite{umejei2025artificial}. Participants expressed varied positions, ranging from optimism regarding efficiency gains to concerns about credibility, misinformation, and institutional readiness. While the study contributed regional perspectives to the literature, it did not provide longitudinal tracking of adoption trends over time.

In Portugal, Canavilhas (2022) surveyed 32 sports editors and found that AI was primarily used for low-complexity tasks~\cite{canavilhas2022artificial}. Participants noted time savings but identified financial limitations and ethical uncertainty as barriers to further implementation. The anonymous nature of the survey precluded follow-up inquiry.

A related content analysis by Canavilhas et al. (2024) reviewed 60 articles from Brazil and Portugal that addressed AI in journalism~\cite{Canavilhas2024Artificial}. Most coverage emphasized innovation and productivity, with limited discussion of employment impacts or ethical concerns. The sampling method, which relied on Google News’s relevance algorithm, may have introduced selection bias.

De-Lima-Santos and Ceron (2021) reviewed 93 AI-related journalism projects recorded in the JournalismAI database~\cite{de-Lima-Santos2021Artificial}. The majority of initiatives involved machine learning and computer vision and were located in North America and Europe. Many received support from technology companies, raising questions about the influence of corporate interests on newsroom innovation. The dataset was not exhaustive, and the study did not assess implementation outcomes~\cite{de-Lima-Santos2021From}.

Albizu-Rivas et al. (2024) interviewed 21 journalists engaged in slow journalism in Spain~\cite{albizu2024artificial}. Respondents reported limited use of AI beyond transcription or editing tools, citing concerns about narrative integrity and creative authorship. The study provided insight into newsroom resistance, though its national scope restricts its comparative significance.

Vergeer (2020) analyzed over 4,200 newspaper articles from the Netherlands between 2000 and 2018~\cite{vergeer2020artificial}. The study observed an increase in AI-related coverage after 2014 but found that topics such as robot-generated journalism remained infrequent. The study employed correlational methods, and multicollinearity was noted as a limitation in model interpretation.

Collectively, these studies indicate that AI adoption in journalism involves more than technological implementation. It intersects with editorial structures, labor practices, and institutional norms. While newsrooms use AI to support production processes, they continue to negotiate its implications for journalistic autonomy, professional roles, and organizational accountability. Future research should examine these dimensions through cross-national comparisons, mixed-method designs, and longitudinal analysis to assess the evolving integration of AI into journalistic systems.

\medskip

Table~\ref{tab:newsroom_summary} summarizes some of the referenced literature that uses various empirical methods to assess newsroom adoption and organizational challenges surrounding AI in journalism.

\begin{table}[htbp]
\centering
\caption{Summary of studies on newsroom AI adoption and organizational challenges.}
\label{tab:newsroom_summary}
\resizebox{\textwidth}{!}{%
\begin{tabular}{@{}p{3.3cm}p{3cm}p{3.3cm}p{3.5cm}p{3.5cm}@{}}
\toprule
\textbf{Reference} & \textbf{Method} & \textbf{Application} & \textbf{Advantages} & \textbf{Limitations} \\
\midrule
Møller et al.\ (2024)~\cite{Moller2024Little} & Semi-structured interviews & Impact of AI on Danish journalism & Redefining expertise, preserving human values & Limited to Denmark, partial AI knowledge \\
Banafi (2024)~\cite{banafi2024review} & Literature review & AI's role in automating journalism & Efficiency tools, ethical AI guidelines & Broad, non-systematic approach \\
Parratt-Fernández et al.\ (2024)~\cite{Parratt-Fernandez2024Spanish} & Case study & Narrativa OÜ and AI integration & Contextual adoption insights & Single-case focus, no large-scale content analysis \\
Noain-Sánchez (2022)~\cite{Noain-Sanchez2022Addressing} & In-depth interviews & AI’s impact across countries & Efficiency gains, ethical awareness & Interpretive nature, limited sample \\
Umejei et al.\ (2025)~\cite{umejei2025artificial} & Semi-structured interviews & AI in African journalism & Diverse perspectives, practical use cases & Lack of policy framework, generalizability \\
Canavilhas (2022)~\cite{canavilhas2022artificial} & Survey & AI in Portuguese sports media & Identifies key barriers & Anonymity limits follow-up \\
Canavilhas et al.\ (2024)~\cite{Canavilhas2024Artificial} & Sentiment analysis & Coverage in Brazil and Portugal & Shows media optimism & News algorithm bias \\
de-Lima-Santos \& Ceron (2021)~\cite{de-Lima-Santos2021Artificial} & Case analysis & Global newsroom adoption & Identifies dominant AI subfields & Non-exhaustive database \\
Albizu-Rivas et al.\ (2024)~\cite{albizu2024artificial} & Interviews & AI in slow journalism & Real-world skepticism of AI & National scope, small sample \\
Vergeer (2020)~\cite{vergeer2020artificial} & Longitudinal analysis & Dutch press AI coverage & Topic trends by outlet type & Multicollinearity, scope gaps \\
\bottomrule
\end{tabular}%
}
\end{table}

\subsection{AI Tools in Journalism: Practice and Innovation}

Recent empirical research documents how AI tools have been implemented in journalistic practice to support automation, summarization, classification, and content generation. Studies vary in scope, method, and focus, yet collectively reflect a growing interest in how these tools reshape editorial processes and raise questions about bias, authorship, and reliability.

Castillo-Campos et al. (2024) applied a mixed-methods experimental design to evaluate bias in news summaries generated by GPT-3.5, GPT-4, and Bing across 199 Spanish headlines~\cite{Castillo-Campos2024Artificial}. The analysis found that GPT-3.5 produced outputs with higher levels of bias compared to GPT-4, which aligned more closely with expert-written summaries. The study emphasized that implicit bias could emerge even in the absence of overtly biased prompts. However, the brevity of the source material and the pace of model development constrained the applicability of the findings.

Ufarte Ruiz and Manfredi Sánchez (2019) examined the Spanish platform Narrativa Inteligencia Artificial through interviews, observation, and a journalist survey~\cite{UfarteRuiz2019Algorithms}. The Gabriele software supported routine content generation, and participants evaluated its outputs as neutral and coherent. At the same time, concerns emerged about lack of interpretive nuance, contextual understanding, and textual diversity. These results were limited by the study’s regional scope and absence of comparative benchmarks.

De-Lima-Santos and Salaverría (2021) used hybrid ethnographic methods to study computer vision applications at La Nación in Argentina~\cite{de-Lima-Santos2021From}. The study showed that although specific tasks benefited from automation, infrastructure constraints and satellite imagery costs restricted broader application. This case reflected disparities in access to AI technologies in the Global South and suggested that adoption depends not only on tool availability but also on resource capacity.

Legal issues surrounding AI-generated journalism were explored by Trapova and Mezei (2022), who combined doctrinal legal analysis with a review of ten European NLG providers~\cite{Trapova2022Robojournalism}. The study concluded that most AI-generated texts do not meet originality requirements for copyright protection under current EU law. The authors emphasized the legal necessity of retaining human authorship standards but did not include data from proprietary newsroom practices.

Calvo Rubio et al. (2024) developed a proof-of-concept system for RTVE, Spain’s public broadcaster, to generate election-related reports~\cite{CalvoRubio2024Methodological}. The study documented improvements in the clarity and structure of AI-generated texts over multiple iterations, though it did not provide metrics on audience reception or effectiveness relative to human-authored reports.

Quinonez and Meij (2024) provided a technical overview of Bloomberg’s implementation of AI in its newsroom, including tools like BloombergGPT and News Innovation Lab projects~\cite{Quinonez2024New}. The analysis described procedures for human oversight and workflow integration. Identified concerns included vulnerability to adversarial attacks and lack of independent validation by external audiences.

Tsourma et al. (2021) introduced EarthPress, a Web 3.0 platform that combined earth observation data, social media monitoring, and automated news generation~\cite{Tsourma2021AI-enabled}. Developed through a participatory workshop involving 72 participants, the system aimed to assist with real-time crisis reporting. Although the platform incorporated diverse data streams, its accuracy and reliability depended heavily on EO data quality and its misinformation detection algorithm remained under-tested.

Demirci and Sagiroglu (2022) created TwitterBulletin, a deep learning-based classifier trained on 35,000 Turkish-language tweets across seven news categories~\cite{Demirci2022Twitterbulletin}. Using a convolutional neural network, the system achieved high classification accuracy, suggesting potential for real-time content categorization and dissemination. However, its performance was not evaluated in live newsroom settings.

Rojas Torrijos and Toural Bran (2019) evaluated AnaFut, a sports bot developed by El Confidencial, through content analysis and expert interviews~\cite{RojasTorrijos2019Automated}. The system enabled automated production of match reports, expanding coverage while reducing human effort. Nonetheless, human oversight remained necessary for improving narrative richness and editorial standards.

Chu and Liu (2024) conducted experimental tests to assess the persuasive capacity of ChatGPT-generated narratives~\cite{Chu2024Can}. Results showed that while machine-written stories were coherent and persuasive, readers exhibited greater skepticism when they were aware of the AI authorship. The study identified credibility as a key challenge for AI-generated narratives, particularly in contexts involving political or emotive content.

Together, these studies indicate that AI tools offer measurable benefits in automation, summarization, and classification, but also introduce challenges related to content bias, legal authorship, credibility, and interpretive depth. The responsible use of AI in journalism requires editorial oversight, legal clarity, and attention to how technologies perform across different socio-technical contexts.

\medskip

Table~\ref{tab:ai_tools_summary} summarizes selected studies examining AI tools in journalism, focusing on practical implementations, their advantages, and inherent limitations.

\begin{table}[htbp]
\centering
\caption{Summary of studies on AI tools in journalism practice and innovation}
\label{tab:ai_tools_summary}
\resizebox{\textwidth}{!}{%
\begin{tabular}{@{}p{3.5cm}p{3.1cm}p{3.4cm}p{3.6cm}p{3.6cm}@{}}
\toprule
\textbf{Reference} & \textbf{Method} & \textbf{Application} & \textbf{Advantages} & \textbf{Limitations} \\
\midrule
Castillo-Campos et al.\ (2024)~\cite{Castillo-Campos2024Artificial} & Mixed-methods experiment & Bias in AI-generated summaries & GPT-4 shows lower bias; expert alignment & Short texts; AI evolves quickly \\
Ufarte Ruiz \& Manfredi Sánchez (2019)~\cite{Parratt-Fernandez2024Spanish} & Mixed-methods over 6 months & AI-generated content analysis (Narrativa IA) & High productivity; basic news automation & Not generalizable; lacks international scope \\
de-Lima-Santos \& Salaverría (2021)~\cite{de-Lima-Santos2021From} & Hybrid ethnography & Computer vision in Argentina’s \textit{La Nación} & CV enhances investigative reporting & Single newsroom focus; infrastructure barriers \\
Trapova \& Mezei (2022) & Legal doctrinal analysis & NLG and copyright in EU & Clarifies legal authorship boundaries & No access to contracts; newsroom opacity \\
Calvo Rubio et al.\ (2024)~\cite{CalvoRubio2024Methodological} & Proof-of-concept with RTVE & Evaluating AI-generated election reports & Improved clarity, coherence & No audience reception; vague quality metrics \\
Quinonez \& Meij (2024)~\cite{Quinonez2024New} & Position paper & Bloomberg’s AI pipeline & Controlled, multimodal, transparent AI journalism & Model errors; limited public impact testing \\
Tsourma et al.\ (2021) & Workshop + system design & EarthPress automation platform & Multimodal automation, disaster reporting & EO data quality, misinformation detection challenges \\
Demirci \& Sagiroglu (2022)~\cite{Demirci2022Twitterbulletin} & AI tool with CNN classification & TwitterBulletin topic categorization & 98\% accuracy; real-time classification & None reported in the summary \\
Rojas Torrijos \& Toural Bran (2019)~\cite{RojasTorrijos2019Automated} & Case study + content analysis & AnaFut sports bot by \textit{El Confidencial} & Expanded match coverage & Repetitive structure; low complexity \\
Chu \& Liu (2024)~\cite{Chu2024Can} & Experiments & Narrative persuasion by ChatGPT & Coherent storytelling; similar to humans & Labeling reduces trust; misinformation risk \\
\bottomrule
\end{tabular}%
}
\end{table}
\subsection{AI, Misinformation, and Verification: Scholarly Developments}

The integration of artificial intelligence (AI) into journalism has altered both the creation and mitigation of misinformation. Researchers have examined how AI tools contribute to or counteract the spread of fake news, while also highlighting the ethical, legal, and perceptual implications of these technologies.

Forja-Peña et al. (2024) analyzed national journalistic ethics codes across Europe using a Delphi method and content analysis~\cite{Forja-Pena2024Ethical}. Their findings showed that most codes lacked explicit references to AI-related responsibilities. Although reliance on AI tools is growing, regulatory frameworks have not evolved at the same pace. The authors recommended revisions addressing transparency, accountability, and algorithmic bias. However, the analysis did not include regional or local codes, limiting contextual comprehensiveness.

Flores Vivar (2019) applied a triangulated research design to assess the utility of AI bots in verifying information~\cite{FloresVivar2019Artificial}. The results indicated that although bots improved accuracy and reliability, they remained slower and less pervasive than the spread of disinformation. The study recommended further behavioral research to investigate user susceptibility and to optimize AI systems for intervention.

Túñez-López et al. (2019) conducted a bibliographic review of literature on AI-driven news automation~\cite{TuneZ-Lopez2019Automation}. Their findings emphasized that AI tools support productivity in factual reporting but may constrain journalistic depth and interpretive analysis, especially in opinion-based formats. They suggested further genre-specific research to address these limitations.

Musi et al. (2024) introduced Botlitica, a GPT-3-based tool designed to identify political propaganda~\cite{Musi2024Botlitica}. In focus group sessions with 18 journalists, participants reported that the tool encouraged critical evaluation of rhetorical tactics. Concerns over algorithmic transparency and user trust persisted, underscoring the need for participatory design in future development.

Namani et al. (2025) developed DeepGuard, a multi-layer AI framework for identifying manipulated images produced by generative models~\cite{Namani2025DeepGuard}. The system achieved high accuracy (up to 99.87\%) but encountered difficulties distinguishing between synthetic and authentic human facial images. The authors recommended integrating multimodal detection systems to improve robustness.

Kim and Desaire (2024) trained a machine learning model to differentiate between human-written and AI-generated student news articles using 13 linguistic features~\cite{Kim2024Detecting}. Their model achieved 98\% accuracy, outperforming existing tools. However, the study tested a narrow set of prompts, which restricts the generalizability of results.

Barredo Ibañez et al. (2023) conducted a systematic review of AI, journalism, and misinformation in the Chinese media landscape~\cite{BarredoIbanez2023Disinformation}. Their study identified the role of AI in reinforcing ideological echo chambers and noted a lack of empirical data due to censorship constraints.

Ali et al. (2024) employed the Unified Theory of Acceptance and Use of Technology 2 (UTAUT2) framework to analyze generative AI adoption among media professionals in the Gulf region~\cite{Ali2024Generative}. They found that hedonic motivation and institutional trust significantly predicted AI use. While culturally informative, the study's regional focus limits broader applicability.

Cazzamatta and Sarısakaloğlu (2025) examined 3,154 verification articles to compare AI deployment in fact-checking organizations across Europe and Latin America~\cite{Cazzamatta2025Mapping}. They reported that European outlets developed proprietary AI systems, while Latin American organizations had lower institutional capacity. The authors recommended expanding research into underrepresented regions and addressing disparities in infrastructure.

Hausken (2024) offered a theoretical framework to distinguish AI-generated images from traditional photojournalism~\cite{Hausken2024Photorealism}. He proposed that researchers differentiate photorealism from photography and assess AI visuals through genre-specific criteria. While the model provides conceptual clarity, it lacks empirical validation and user testing.

Gutiérrez-Caneda and Vázquez-Herrero (2024) analyzed ten fact-checking organizations through interviews and case studies~\cite{Gutierrez-Caneda2024Redrawing}. Their analysis indicated that AI enhanced the efficiency of early-stage misinformation detection. Nonetheless, financial limitations and platform-related constraints persisted, and growing skepticism toward AI-based verification among users was noted.

García-Faroldi et al. (2025) surveyed 1,550 Andalusian citizens to assess perceptions of AI's role in misinformation~\cite{Garcia-Faroldi2025Unmasking}. Respondents generally believed that AI facilitates fake news creation, but they also recognized its potential to detect and counteract disinformation. Attitudes varied significantly by gender, education, socioeconomic status, and political orientation. The authors recommended targeted media literacy initiatives.

Collectively, these studies emphasize that AI's influence on misinformation is multifaceted. While AI contributes tools for verification and detection, it also introduces challenges in ethics, user trust, and regulatory adequacy. Future research should evaluate these tensions systematically and explore the implications for policy and media literacy across diverse sociopolitical contexts.

\medskip

Table~\ref{tab:fake_news_summary} summarizes selected studies addressing AI’s role in fake news detection, verification, and public engagement, highlighting diverse methods, contexts, and implications.

\begin{table}[htbp]
\centering
\caption{Summary of studies addressing AI and fake news in journalism}
\label{tab:fake_news_summary}
\resizebox{\textwidth}{!}{%
\begin{tabular}{@{}p{3.8cm}p{3.1cm}p{3.4cm}p{3.8cm}p{3.8cm}@{}}
\toprule
\textbf{Reference} & \textbf{Method} & \textbf{Application} & \textbf{Advantages} & \textbf{Limitations} \\
\midrule
Forja-Pena et al.\ (2024)~\cite{Forja-Pena2024Ethical} & Delphi + content analysis & Ethics codes \& AI coverage & Reveals ethical gaps in AI mention & Limited to European codes \\
Flores Vivar (2019)~\cite{FloresVivar2019Artificial} & Triangulation (qual + quant) & AI bots vs. disinformation & Verifies info; enhances credibility & Bots' scale; human judgment impact \\
Túñez-López et al.\ (2019)~\cite{TuneZ-Lopez2019Automation} & Comparative review & News automation & Productivity gains; trend analysis & Exploratory; lacks genre depth \\
Musi et al.\ (2024)~\cite{Musi2024Botlitica} & Focus group + tool deployment & Botlitica for propaganda & Boosts critical skills & Trust concerns; usability issues \\
Namani et al.\ (2025)~\cite{Namani2025DeepGuard} & Ensemble classification model & Fake image detection & High accuracy; source attribution & Misclassifies faces; dataset limits \\
Kim \& Desaire (2024)~\cite{Kim2024Detecting} & ML classification & ChatGPT text detection & High detection accuracy & Narrow prompt range \\
Barredo Ibañez et al.\ (2023)~\cite{BarredoIbanez2023Disinformation} & Systematic review & AI, disinfo, and Chinese journalism & Identifies echo chamber risks & Censorship context; lacks empirical data \\
Ali et al.\ (2024)~\cite{Ali2024Generative} & UTAUT2 + survey & GenAI adoption in Arabian Gulf & Hedonic motivation, trust insights & Regional focus; cultural scope \\
Cazzamatta \& Sarısakaloğlu (2025)~\cite{Cazzamatta2025Mapping} & Lit. review + content analysis & Fact-checking tools across regions & Identifies tool disparity by region & Needs Global South inclusion \\
Hausken (2024)~\cite{Hausken2024Photorealism} & Theoretical analysis & AI image ethics & Clear conceptual frameworks & No user validation \\
Gutiérrez-Caneda \& Vázquez-Herrero (2024)~\cite{Gutierrez-Caneda2024Ethics} & Case studies + interviews & AI in fact-checking & Improves efficiency \& early detection & Platform limitations; public distrust \\
García-Faroldi et al.\ (2025)~\cite{Garcia-Faroldi2025Unmasking} & Survey (n=1550) & Public perception of AI \& fake news & Links AI to both problem \& solution & Region-specific; influenced by education \\
\bottomrule
\end{tabular}%
}
\end{table}
\subsection{Ethical and Professional Considerations}

The integration of artificial intelligence (AI) into journalism introduces a complex spectrum of ethical, legal, and professional challenges. Scholars have responded by developing frameworks, evaluating regulatory needs, analyzing newsroom practices, and investigating public perceptions.

Dierickx et al. (2024) developed the Accuracy-Fairness-Transparency (AFT) framework to assess data quality in AI-driven journalism~\cite{Dierickx2024DataCentric}. Their findings suggest that prioritizing data quality over dataset size enhances AI reliability. However, they acknowledged that the framework's effectiveness diminishes when applied to large foundational models. The authors emphasized that journalist involvement and improved data literacy are essential for responsible AI development.

Molitorisz (2024) examined algorithmic news curation through a regulatory lens~\cite{molitorisz2024,Molitorisz2024Legal}. He found that opaque digital systems diminish democratic autonomy by overwhelming users with content while shielding decision-making processes. Based on Australian and European regulatory comparisons, he advocated for algorithmic regulators and participatory design to ensure digital platforms uphold journalistic integrity and democratic values.

Using survey and experimental data, Piasecki et al. (2024) explored how readers respond to AI-generated news~\cite{Piasecki2024AIGenerated}. Participants felt manipulated when news content lacked disclosure of AI authorship. The study showed that basic labels under the EU AI Act (Article 50) fail to enhance transparency or build trust. The authors recommended more substantive forms of transparency that empower audiences to critically evaluate AI-generated journalism.

Lukina et al. (2022) addressed ethical standards for AI in Russian media~\cite{Lukina2022Artificial}. They proposed that journalists and developers share responsibility for content accuracy, that audiences be informed of AI involvement, and that machines not assume moral decision-making roles. They called for codified ethical provisions to keep pace with technological developments in media.

Legal studies by Díaz-Noci (2020) and Kuai (2024) interrogated copyright regimes~\cite{Diaz-Noci2020Artificial,Kuai2024Unravelling}. Díaz-Noci argued that only content with human involvement deserves full copyright protection, proposing shorter terms for AI-generated works to promote public access. Kuai, in a comparative study of China, the U.S., and the EU, found that current copyright frameworks favor tech companies, threatening journalistic independence and failing to reflect the complexities of automated authorship.

Simon (2022, 2023) introduced the concept of “infrastructure capture” to describe how reliance on proprietary AI tools erodes editorial autonomy~\cite{Simon2022Uneasy,Simon2023Escape}. Through interviews with journalists and technologists, he documented how news organizations lose control over content creation and distribution, emphasizing the need for in-house AI development. Spyridou and Ioannou (2025) similarly warned that AI hype conceals structural dependencies and risks undermining journalism’s civic mission unless embedded within human-centered editorial frameworks~\cite{Spyridou2025Exploring}.

Empirical studies on newsroom practices reflect diverse approaches to AI ethics. Becker et al. (2025) reviewed 52 editorial guidelines and found common principles such as transparency, human oversight, and accountability~\cite{Becker2025Policies}. However, they noted limited enforcement and significant regional disparities. 

Sánchez-García et al. (2025) found that newsrooms across Europe and Latin America are crafting self-regulatory norms, though many remain underdeveloped or poorly institutionalized~\cite{Sanchez-Garcia2025Media}.

Several studies highlighted the limitations of AI-generated content in fulfilling journalism’s ethical commitments. Breazu and Katsos (2024) analyzed narratives produced by ChatGPT-4 and found that although the model avoids harmful stereotypes, it lacks the human intent central to responsible journalism~\cite{Breazu2024ChatGPT-4}. 

González-Arias et al. (2024) showed that while AI can mimic subjectivity, it struggles to adapt tone and voice to context, which risks misrepresenting complex stories~\cite{Gonzalez-Arias2024Anthropomorphic}.

Research on public attitudes indicates that trust in AI journalism depends on perceived human oversight. Heim and Chan-Olmsted (2023) reported that consumers preferred hybrid human–AI models over full automation~\cite{Heim2023Consumer}
.

Lermann Henestrosa and Kimmerle (2024) found consistent support for human authorship and noted widespread uncertainty about how AI-generated content works, underlining the need for public education and transparent labeling~\cite{LermannHenestrosa2024Understanding}.

Ethical risks remain a pressing concern. Gutiérrez-Caneda et al. (2024) documented worries about data privacy, bias, and professional deskilling among journalists~\cite{Gutierrez-Caneda2024Ethics}. Al-Zoubi et al. (2024) found similar concerns in Jordanian newsrooms, where limited regulation and training constrain ethical deployment~\cite{Al-Zoubi2024Artificial}. Matich et al. (2025) and Wu et al. (2022) investigated AI’s impact on visual journalism and recommendation systems, respectively, concluding that misinformation, audience manipulation, and labor displacement pose major risks~\cite{matich2025,Wu2022Removing}.

Borden et al. (2024) responded by promoting a Human-Centered AI framework, emphasizing that ethical responsibility must remain with human actors~\cite{Borden2024Introduction}. They argued that transparency, especially through clear AI use disclosures, is fundamental to preserving public trust and journalistic integrity.

Together, these studies reveal that ethical and professional issues in AI journalism are deeply intertwined with legal structures, editorial norms, technological dependencies, and public expectations. Addressing these challenges requires interdisciplinary collaboration, updated regulatory tools, and a continued commitment to human oversight and accountability.

Table~\ref{tab:ethical_summary} summarizes the referenced studies with respect to their methods, applications, identified advantages, and limitations.

\begin{small}
\begin{longtable}{p{.12\linewidth} p{.15\linewidth} p{.18\linewidth} p{.2\linewidth} p{.28\linewidth}}
\caption{Summary of Ethical and Professional Considerations in AI-driven Journalism. Only studies cited in the narrative are included.}
\label{tab:ethical_summary_trimmed} \\
\toprule
\textbf{Reference} & \textbf{Method} & \textbf{Application} & \textbf{Advantages} & \textbf{Limitations} \\
\midrule
\endfirsthead

\caption[]{(Continued) Summary of Ethical and Professional Considerations in AI-driven Journalism.} \\
\toprule
\textbf{Reference} & \textbf{Method} & \textbf{Application} & \textbf{Advantages} & \textbf{Limitations} \\
\midrule
\endhead

\midrule
\endfoot
\bottomrule
\endlastfoot

Dierickx et al. (2024)~\cite{Dierickx2024DataCentric} & Interdisciplinary framework (AFT) & Data quality ethics in journalism AI & Emphasizes data quality over volume; promotes ethical literacy & Hard to apply to foundational LLMs \\
Molitorisz (2024)~\cite{molitorisz2024} & Regulatory analysis & Algorithmic regulation & Advocates algorithmic regulators and participatory design & Needs broader implementation evidence \\
Piasecki et al. (2024)~\cite{Piasecki2024AIGenerated} & Survey + experimental design & Transparency in AI-generated news & Shows basic disclosure fails; promotes richer transparency & Short-term study; EU-specific framing \\
Lukina et al. (2022)~\cite{Lukina2022Artificial} & Case study of Russian news outlets & Ethical standards in AI journalism & Proposes joint responsibility between journalists and developers & Region-specific; lacks enforcement mechanisms \\
Díaz-Noci (2020)~\cite{Diaz-Noci2020Artificial} & Legal review & Copyright protection in AI journalism & Supports reduced protection for AI-only content & Jurisdictional inconsistency \\
Kuai (2024)~\cite{Kuai2024Unravelling} & Comparative legal study & IP and copyright across U.S., EU, and China & Reveals imbalance favoring tech firms & Legal focus; little newsroom perspective \\
Simon (2022)~\cite{Simon2022Uneasy} & Theoretical analysis & Infrastructure capture in AI workflows & Conceptualizes platform-induced dependency & No empirical evidence provided \\
Simon (2023)~\cite{Simon2023Escape} & Interviews with technologists & Proprietary platform reliance & Highlights erosion of editorial autonomy & Western-focused; tech-centric sample \\
Spyridou and Ioannou (2025)~\cite{Spyridou2025Exploring} & Political economy perspective & Human–AI editorial collaboration & Advocates civic-centered AI integration & Conceptual emphasis; no empirical test \\
Becker et al. (2025)~\cite{Becker2025Policies} & Review of 52 guidelines & AI ethics in editorial policy & Finds consensus on transparency, oversight & Weak enforcement; regional variation \\
Sánchez-García et al. (2025)~\cite{Sanchez-Garcia2025Media} & Content analysis & Emerging self-regulation norms & Captures diverse practices in Europe and Latin America & Many guidelines are underdeveloped \\
Breazu and Katsos (2024)~\cite{Breazu2024ChatGPT-4} & Narrative analysis & Bias in AI-generated migration stories & Shows GPT-4 avoids overt stereotypes & Lacks human contextual nuance \\
González-Arias et al. (2024)~\cite{Gonzalez-Arias2024Anthropomorphic} & Content analysis & Subjectivity in LLM-generated news & Finds AI mimics but cannot adapt tone to context & Risks misrepresentation of complex issues \\
Heim and Chan-Olmsted (2023)~\cite{Heim2023Consumer} & Survey-based study & Trust in AI vs. human news authorship & Readers favor hybrid models & U.S.-based; limited generalizability \\
Lermann Henestrosa and Kimmerle (2024)~\cite{LermannHenestrosa2024Understanding} & Survey + regression analysis & Public perception of AI journalism & Emphasizes role of education and labeling & Awareness gaps; cross-sectional design \\
Gutiérrez-Caneda et al. (2024)~\cite{Gutierrez-Caneda2024Ethics} & Qualitative interviews & AI ethics in European newsrooms & Identifies concerns over deskilling and bias & Small sample; Eurocentric \\
Al-Zoubi et al. (2024)~\cite{Al-Zoubi2024Artificial} & Interviews with Jordanian journalists & AI ethics in low-regulation settings & Documents risks in under-resourced contexts & Regional scope; limited sample \\
Matich et al. (2025)~\cite{matich2025} & Qualitative interviews & Visual journalism and misinformation & Identifies objectivity loss and labor risks & Needs stronger audience-centered data \\
Wu et al. (2022)~\cite{Wu2022Removing} & Computational analysis & Sentiment debiasing in recommender systems & Demonstrates reduced bias with adversarial learning & Narrow focus on sentiment only \\
Borden et al. (2024)~\cite{Borden2024Introduction} & Theoretical/conceptual analysis & Human-centered AI in journalism & Re-centers ethical accountability on humans & Requires operationalization of concepts \\
\end{longtable}
\end{small}

\section{Discussion}\label{dis}
This systematic review synthesizes empirical findings from 72 peer-reviewed studies on artificial intelligence (AI) in journalism, offering insight into the field’s rapid growth, thematic diversity, and structural tensions between technological innovation and normative frameworks. The integration of AI into journalism has shifted from exploratory experimentation to widespread adoption across news production, verification, and distribution~\cite{de-Lima-Santos2021Artificial,UfarteRuiz2019Algorithms,Quinonez2024New}. However, the literature reflects persistent challenges related to ethical ambiguity~\cite{Piasecki2024AIGenerated,molitorisz2024}, geographic concentration~\cite{canavilhas2022artificial}, disciplinary fragmentation~\cite{Dierickx2024DataCentric,Gonzalez-Arias2024Rethinking}, and the reconfiguration of professional roles~\cite{Moller2024Little,albizu2024artificial}.
\subsection{Reframing Journalism Through AI: Evolving Roles, Norms, and Practices}

AI technologies are not simply tools that automate workflows but are actively reshaping what it means to be a journalist. As AI systems increasingly co-author content or influence editorial decisions, traditional concepts of authorship, agency, and accountability are being renegotiated~\cite{gonzalez2024,Trapova2022Robojournalism}. Journalists face new ethical dilemmas in determining when, how, and whether to disclose AI involvement—decisions that directly affect public trust and professional identity~\cite{Borden2024Introduction,Piasecki2024AIGenerated}.

Several studies suggest that journalists are not passive recipients of technological change. Rather, they engage critically with AI systems, often modifying or resisting their use to align with established editorial norms and institutional values~\cite{Moller2024Little,albizu2024artificial}. This supports an agency-oriented view of technological adoption and aligns with theories of technological domestication, which emphasize the processes of adaptation, negotiation, and resistance within organizational settings~\cite{de-Lima-Santos2021Artificial}.

A consistent issue across the literature is the lag between technological implementation and the development of ethical or regulatory frameworks. The rapid deployment of generative models such as GPT-3, ChatGPT, and Gemini has outpaced institutional mechanisms designed to ensure transparency, accuracy, and fairness~\cite{Simon2023Escape,Piasecki2024AIGenerated}. In many cases, practices related to AI-generated content disclosure remain inconsistent and insufficiently standardized~\cite{molitorisz2024,Lukina2022Artificial}.

This reactive posture reinforces the urgency for anticipatory governance—an approach that incorporates ethical considerations during the design, deployment, and evaluation stages of AI systems~\cite{Borden2024Introduction,Dierickx2024DataCentric}. Scholars argue that ethical journalism in the AI era cannot rely solely on retrospective correction but must involve proactive collaboration between journalists, technologists, and policymakers~\cite{Trapova2022Robojournalism}.

The review also identifies a strong geographic imbalance in existing research. Much of the empirical literature is concentrated in high-income countries such as the United States, United Kingdom, Spain, and Germany~\cite{Parratt-Fernandez2024Spanish,de-Lima-Santos2021From}, while the Global South remains underrepresented. Although studies like Umejei et al. (2025) provide important insights from Nigeria, Ghana, Kenya, and South Africa~\cite{umejei2025artificial}, comparable research from South Asia, Latin America, or the Middle East is relatively scarce.

This underrepresentation limits the field’s capacity to account for contextual factors such as censorship, infrastructure disparities, and linguistic diversity~\cite{Kuai2024Unravelling,Canavilhas2024Artificial}. Moreover, the dominance of Euro-American paradigms risks shaping normative standards and innovation paths that are not universally applicable~\cite{Spyridou2025Exploring}. Scholars have called for inclusive research that foregrounds alternative epistemologies and centers local journalistic practices and sociopolitical contexts~\cite{de-Lima-Santos2021Artificial}.

Another major finding is the fragmentation of research across disciplinary boundaries. Studies focused on the technical performance of AI systems; often remain disconnected from those examining ethical values, newsroom dynamics, or public trust~\cite{Demirci2022Twitterbulletin,Castillo-Campos2024Artificial,Gutierrez-Caneda2024Redrawing,Heim2023Consumer}. Similarly, legal inquiries into copyright and authorship rarely intersect with empirical studies on audience reception or platform accountability~\cite{Diaz-Noci2020Artificial,Kuai2024Unravelling}.

This disciplinary siloing has impeded the field’s ability to produce comprehensive, actionable insights. Scholars have therefore advocated for interdisciplinary integration that connects journalism studies, data science, law, and communication ethics~\cite{Cheng2025When,Cools2024Uses}. Mixed-method studies—such as those combining ethnography with computational analysis—offer promising models for bridging this gap~\cite{Bastian2021Safeguarding,Gutierrez-Caneda2024Redrawing}.

Theoretically, the review calls for revisiting key assumptions in journalism studies. Traditional notions of professional autonomy and human authorship must be adapted to account for hybrid human–machine environments. Theories of infrastructural power~\cite{Simon2023Escape}, hybrid agency~\cite{siren2023crossroads}, and human–machine communication~\cite{Partha2024Artificial} provide valuable frameworks for understanding how AI reshapes editorial decision-making and institutional authority.

Practically, the findings emphasize the need for clear newsroom policies on AI use. These include guidelines for attribution, quality control, human oversight, and ethical redress. As AI becomes more integrated into journalistic routines, journalists must be trained in both technical literacy and critical evaluation skills~\cite{canavilhas2022artificial,umejei2025artificial}.

From a policy perspective, current legal frameworks such as the EU AI Act are insufficient for addressing the specific challenges of AI-generated journalism. Scholars argue for more robust transparency requirements that include context, intention, and traceability of AI-generated content~\cite{Piasecki2024AIGenerated,Lukina2022Artificial}. Additionally, public-interest interventions are needed to ensure that media systems in resource-poor environments are not excluded from the benefits of AI innovation~\cite{Kuai2024Unravelling,Sanchez-Garcia2025Media}.

This review is limited to English-language, peer-reviewed studies indexed in Scopus and Web of Science, potentially underrepresenting practitioner-based or non-Western research. It excludes conceptual articles, which may omit key normative insights. No formal quality appraisal of study rigor was conducted. Moreover, the literature reflects developments only up to mid-2025, and emerging innovations in generative AI may present new challenges beyond this scope.

Despite these constraints, the review provides a timely synthesis of a dynamic and expanding research domain, identifying key structural gaps and offering directions for inclusive, interdisciplinary inquiry.
\subsection{Sentiment Reflections on AI in Journalism}
The sentiment analysis conducted on 72 article abstracts offers a reflective window into the underlying affective landscape of AI discourse in journalism. Rather than revealing a binary of optimism versus skepticism, the data presents a complex affective terrain shaped by co-existing excitement and anxiety. The prevalence of positively valenced terms such as \textit{“intelligence,” “support,” “accuracy,” “benefits,”} and \textit{“efficiency”}—collectively accounting for over 60\% of positive mentions—signals a general enthusiasm surrounding AI’s capacity to enhance journalistic workflows, content verification, and innovative storytelling (see Figure~\ref{fig:top_positive_words}). Notably, the prominence of the term \textit{“intelligence”} alone, representing nearly 35\% of the positive frequency share, reflects a discourse that is heavily shaped by technical admiration and the symbolic appeal of automation.

Conversely, the negative lexicon—dominated by terms such as \textit{“bias,” “concern,” “risks,” “misinformation,”} and \textit{“manipulation”—}reveals recurring anxieties around AI’s normative misalignment with journalistic ethics (Figure~\ref{fig:top_negative_words}). These words do not simply indicate disagreement with AI per se, but rather reflect apprehension about its unchecked deployment, the opacity of algorithmic decisions, and the potential erosion of editorial accountability. The frequency and consistency of these concerns across years underscore a growing awareness that technological sophistication does not inherently translate to ethical clarity.

Furthermore, the temporal analysis of sentiment trends demonstrates a subtle but consistent tension: while the overall tone remains cautiously positive, the presence of negative sentiment has not declined—instead, it has grown more specific and targeted (Figure~\ref{fig:sentiment_trend_year}). This suggests that as AI tools become more integrated into newsroom routines, the discursive focus is shifting from broad conceptual excitement to situated ethical dilemmas. This trend is further confirmed by the annual mean sentiment score ($\bar{s}_y$), which remains positive yet shows variation aligned with major shifts in AI adoption and critique (Figure~\ref{fig:yearly_sentiment}). These findings suggest that future discussions on AI in journalism must move beyond general endorsements or rejections and engage with the deeper socio-technical entanglements that shape trust, transparency, and public interest in mediated information systems.

\subsection{Structural Patterns and Gaps from Bibliometric Mapping}
The bibliometric analysis highlights a rapidly expanding but unevenly distributed research landscape on AI in journalism. The sharp growth in publication volume, particularly from 2020 onward, illustrates a shift from speculative engagement to systematic inquiry (see Figure~\ref{fig:pub_trend})—suggesting that AI’s role in journalism has moved from technological possibility to institutional reality. This trajectory is further supported by the concentration of high-impact contributions clustered around key journals (Figure~\ref{fig:top_journals}), reflecting a growing legitimacy of AI topics within mainstream scholarly forums.

Despite this growth, the data reveal persistent structural asymmetries. The collaboration networks are notably dense within Europe and North America, while scholars from the Global South—particularly regions like South Asia, Latin America, and parts of Africa—remain marginal (see Figure~\ref{fig:top_countries}). This geographic imbalance risks reinforcing epistemic hierarchies, where dominant paradigms and normative assumptions are shaped by contexts with comparatively advanced infrastructure and institutional support.

Author co-citation and keyword co-occurrence patterns also suggest thematic clustering around innovation, misinformation, and ethics (Figures~\ref{fig:fig_countries} and~\ref{fig:fig_keywords}). However, these clusters appear relatively siloed, with limited interdisciplinary overlap. This fragmentation may hinder the development of holistic frameworks that can address AI’s multifaceted impact on journalism—from algorithmic design to newsroom dynamics and audience perception. The low density of cross-domain citation ties, especially between technical and normative research domains, reflects an ongoing disconnection that future scholarship must reconcile.

Taken together, the bibliometric findings raise important questions about research equity, interdisciplinarity, and agenda-setting. They point to the need for more globally inclusive, critically engaged, and methodologically diverse collaborations. Bridging these divides is essential not only for theoretical comprehensiveness but also for designing AI applications that are context-sensitive and aligned with journalism’s democratic mission.

\subsection{Cross-Thematic Synthesis and Gaps}
This section identifies three critical gaps in the literature: temporal misalignment, geographic concentration, and disciplinary fragmentation. These gaps carry implications for research design, newsroom practice, and AI governance in journalism.

A recurring pattern across the literature is the time lag between the rapid implementation of AI tools in newsrooms and the slower development of normative, ethical, and regulatory responses~\cite{Simon2023Escape,Piasecki2024AIGenerated}. While tools such as GPT-3, GPT-4, and BloombergGPT are increasingly integrated into journalistic workflows, scholarly reflection on their implications—such as transparency, authorship, and accountability—often arrives much later~\cite{Quinonez2024New,molitorisz2024,Lukina2022Artificial}. This misalignment results in reactive rather than proactive governance, increasing the risk of misinformation, editorial malpractice, and erosion of public trust. Future research must align innovation cycles with anticipatory ethical reflection and institutional design~\cite{Borden2024Introduction,Dierickx2024DataCentric,Trapova2022Robojournalism}.

The bulk of empirical studies on AI in journalism originate from predominantly high-income countries, including the United States, United Kingdom, Spain, Germany, and Australia~\cite{Parratt-Fernandez2024Spanish,de-Lima-Santos2021From}. In contrast, regions in the Global South—particularly South Asia, Sub-Saharan Africa, and Latin America—are underrepresented despite facing distinct challenges such as censorship, infrastructural limitations, linguistic bias, and authoritarian pressure~\cite{umejei2025artificial,Kuai2024Unravelling,Canavilhas2024Artificial}. This geographic imbalance limits the generalizability of findings and fails to account for context-specific variables that influence AI adoption and adaptation~\cite{Spyridou2025Exploring}. More inclusive and comparative research is needed to understand how political, cultural, and infrastructural conditions shape the ethical and practical outcomes of AI journalism worldwide~\cite{de-Lima-Santos2021Artificial}.

Despite the inherently interdisciplinary nature of AI in journalism, research often remains siloed. Technical studies tend to focus on algorithmic accuracy, system design, or bias mitigation without addressing their normative implications~\cite{Demirci2022Twitterbulletin,Castillo-Campos2024Artificial}. Conversely, ethics-oriented and audience-centered research sometimes lacks technological specificity or empirical grounding in actual newsroom practices~\cite{Gutierrez-Caneda2024Redrawing,Heim2023Consumer}. For example, studies exploring legal questions about copyright rarely intersect with those examining audience perceptions or platform design~\cite{Diaz-Noci2020Artificial,Kuai2024Unravelling}. This fragmentation undermines holistic understanding and prevents the development of comprehensive frameworks for responsible AI integration. Future research must bridge these divides by combining computational, legal, sociological, and media studies perspectives~\cite{Cheng2025When,Cools2024Uses}. Mixed-method approaches that incorporate ethnography, experiments, computational modeling, and policy analysis can offer richer insights into how AI systems operate within complex journalistic ecosystems~\cite{Bastian2021Safeguarding,Gutierrez-Caneda2024Redrawing}.

These three intersecting gaps—temporal, geographic, and disciplinary—highlight critical weaknesses in the current scholarly landscape. Addressing them requires structural shifts in how research is conducted, funded, and evaluated. Future work must prioritize cross-regional collaborations, interdisciplinary designs, longitudinal tracking, and ethical co-design processes involving journalists, technologists, and civil society~\cite{Simon2023Escape,siren2023crossroads,Partha2024Artificial}. AI-driven journalism must be understood not only as a technological phenomenon but as a socio-political process embedded in power, labor, and communication systems~\cite{canavilhas2022artificial,umejei2025artificial,Piasecki2024AIGenerated,Lukina2022Artificial,Kuai2024Unravelling,Sanchez-Garcia2025Media}.
\subsection{Limitations and Future Directions}
While this review offers a broad and updated synthesis of the AI–journalism intersection, several limitations remain that may constrain the generalizability and completeness of the findings.

\textbf{Limitations:}
\textbf{Limitations:}
\begin{itemize}
    \item Although the initial search spanned articles from 2010 to mid-2025, the inclusion and exclusion criteria—such as topic relevance, document type, and minimum citation threshold—were consistently satisfied only from 2016 onward. As a result, the dataset and corresponding analysis primarily reflect the period 2016–2025, limiting insights into any earlier foundational works.
    \item This review only includes peer-reviewed articles written in English and indexed in Scopus and Web of Science. As a result, grey literature, non-indexed regional work, and practitioner-based contributions remain excluded. This may reinforce Western academic dominance and overlook valuable insights from underrepresented contexts.
    \item The thematic classification relies on keyword co-occurrence and abstract-level content. While this offers high-level trends, it may underrepresent nuance found in full-text discussions or ethnographic detail.
    \item Given the rapid pace of AI innovation, especially in generative models, some very recent developments or conference proceedings may not yet be indexed and thus are not reflected in the current analysis.
\end{itemize}

\textbf{Future Research Directions:}
\begin{itemize}
    \item Future studies should explore AI adoption in journalistic settings across the Global South, particularly in regions facing censorship, infrastructure challenges, or unique linguistic environments.
    \item Mixed-method approaches that integrate computational models with newsroom ethnography, interviews, or policy analysis are needed to bridge the gap between technical design and socio-cultural impact.
    \item Research should give attention to how AI tools are being co-opted, resisted, or adapted by journalists on the ground, moving beyond deterministic or celebratory accounts of technological progress.
    \item There is also a need for longitudinal studies that track evolving newsroom policies, audience reactions, and algorithmic performance over time, particularly in response to emerging regulations like the EU AI Act.
\end{itemize}

\section{Conclusion}\label{con}
This systematic review of 72 peer-reviewed empirical studies (2010–mid-2025) shows that artificial intelligence (AI) is transforming journalism beyond automation—reshaping professional roles, ethical standards, and newsroom routines. Across the domains of newsroom practice, misinformation, innovation, and ethics, AI technologies are influencing how journalism is produced, verified, and received.

Three persistent structural issues remain: a lag between technological deployment and ethical response; a concentration of research in high-income contexts; and siloed disciplinary inquiry. Addressing these challenges requires cross-cultural, interdisciplinary, and participatory approaches to research and policy design.

Nevertheless, the field is advancing. Journalists are negotiating AI integration with skepticism and agency. Ethical concerns are gaining traction in policy debates. Mixed-methods research is offering deeper insight into AI's sociotechnical dynamics. As AI’s role in journalism grows, the field must embrace inclusive governance, human-centered design, and shared responsibility to align technological advancement with the democratic mission of journalism.

\section*{Conflict of interest}
The authors declare no conflict of interest.

\bibliographystyle{unsrt}  
\bibliography{main}

\end{document}